\documentclass[12pt]{article}
\usepackage{amsmath,amsfonts, epsf,epsfig,color,latexsym}
\usepackage{bbold}
\numberwithin{equation}{section}

\ifx\ltimes\undefined
\newcommand{\ltimes}{{\kern3pt\hbox{\vrule width 0.4pt height 5.30pt
depth .0pt}\kern-1.76pt\times\kern1pt}} \fi

\ifx\lrtimes\undefined
\newcommand{\rtimes}{{\kern1pt\times\kern-4.76pt\kern3pt\hbox{\vrule width 0.4pt height 5.30pt
depth .0pt}}} \fi

\def\Z {\mathbb{Z}}
\def\R {\mathbb{R}}

\def\XX {\mathbb{X}}

\def\ti{\tilde}

\def\d{\text{d}}

\def\G{\Gamma}

\def\cG{{\cal G}}
\def\cX{{\cal X}}

\begin{document}

\begin{titlepage}

\hfill QMUL-PH-09-08

\vspace*{30mm}

\begin{center}

\textbf{\Large Flux compactifications, twisted tori\\ and doubled geometry} \\

\vspace*{20mm}

{ R A Reid-Edwards} \\
\vspace*{7mm}

{\em Centre for Research in String Theory }\\
{\em Queen Mary, University of London } \\ {\em Mile End Road} \\
{\em London, E1 4NS, U.K.} \\

\vspace*{15mm}
\vspace*{15mm}

\end{center}

\begin{abstract}
\noindent In \cite{New} an $O(D,D)$-covariant sigma model describing the embedding of a closed world-sheet into the $2D$-dimensional twisted torus $\cX$ was proposed. Such sigma models provide a universal description of string theory with target spaces related by the action of T-duality. In this article a six-dimensional toy example is studied in detail. Different polarisations of the six-dimensional target space give different three-dimensional string backgrounds including a nilmanifold with $H$-flux, a T-fold with $R$-flux and a new class of T-folds. Global issues and connections with the doubled torus formalism are discussed. Finally, the sigma model introduced in \cite{New}, describing the embedding of a world-sheet into $\cX$, is generalised to one describing a target space which is a bundle of $\cX$ over a base $M_d$, allowing for a more complete description of the associated gauged supergravity from the world-sheet perspective to be given.
\end{abstract}

\vfill

\noindent {rreidedwards@gmail.com}\\(Visiting researcher)

\end{titlepage}

\newpage

\section{Introduction}

The duality symmetries of string/M-theory are discrete (gauge) symmetries that do not preserve the distinction between metric and non-metric degrees of freedom \cite{Hull:1994ys}. Indeed, the dualities indicate that the degrees of freedom of the theory can be repackaged in many
different ways which lead to superficially different descriptions of the same underlying physics. The perspective advocated in this paper  is predicated on the idea
that a fundamental formulation of string/M-theory theory should exist in which the T- and U-duality symmetries are manifest (see \cite{Hull Duality and strings space and time, Hull ``A geometry for non-geometric string backgrounds''} for earlier statements of this idea).

A key feature of such a description is expected to be a Stringy Relativity Principle, in which the fact that the distinction between
space and time is dependent on ones frame of reference, is generalized to a principle in which the distinction between the metric and other fields in the theory is dependent on ones
perspective, or duality frame. Just as the natural framework for relativity generalizes three-dimensional Newtonian space to
four-dimensional space-time, the natural framework for M-Theory would generalize (ten- or eleven-dimensional) space-time into a higher-dimensional geometry in
which auxiliary dimensions would be related to non-metric degrees of freedom; just as the Lorenz transformations act naturally on space-time,
the duality symmetries of string- and M-theory would be discrete geometric symmetries of this generalized space. If this perspective is correct, then it is anticipated that there exist many string theory backgrounds that cannot be understood in terms of conventional notions of space-time but would find a natural formulation in terms of such a manifestly duality-covariant formalism. In particular, there is evidence that many massive, gauged supergravities cannot be naturally embedded in string theory without such a framework \cite{Shelton ``Nongeometric flux compactifications'',Hull ``Flux compactifications of string theory on twisted tori'',Samtleben:2008pe,Dall'Agata:2007sr}.

\subsection{Doubled geometry}

A realization of such a duality-covariant formalism has recently been constructed in the form of a doubled geometry \cite{New,Hull ``A geometry for non-geometric string backgrounds''}, where the T-duality group is realized as a subgroup of the action of large
diffeomorphisms on a higher-dimensional `doubled geometry'. For example, strict T-duality \cite{Buscher ``A Symmetry of the String Background Field Equations''} identifies string theory on a circle $S^1$ of radius $r$ with a string theory on
a circle $\widetilde{S}^1$ of radius $r^{-1}$ (in appropriate units) as dual descriptions of the same physics. In this case, the doubled geometry is the torus $T^2=S^1\times
\widetilde{S}^1$ and the conventional space is recovered as the quotient $S^1=T^2/\widetilde{S}^1$, whereas the dual space is recovered as
$\widetilde{S}^1=T^2/S^1$. The doubled geometry $T^2=S^1\times
\widetilde{S}^1$ encodes information on the space $S^1$ and its T-dual $\widetilde{S}^1$ in a manifestly T-duality symmetric way. More generally, one can consider a conventional $n$-dimensional torus fibration over a non-contractible, $d$-dimensional base $M_d$ with transition functions in $SL(n;\Z)$ - the mapping class group of the $T^n$ fibres. The T-duality group here is $O(n,n;\Z)$ and the doubled space is then a $(2n+d)$-dimensional `doubled torus' bundle ${\cal T}$
\begin{eqnarray}
T^{2n}\hookrightarrow & {\cal T} &\nonumber\\
&\downarrow&\nonumber\\
&M_d&\nonumber
\end{eqnarray}
with base $M_d$, $2n$-dimensional fibres $T^{2n}\simeq T^n\times\widetilde{T}^n$, where $\widetilde{T}^n$ is given by performing strict T-dualities along all $n$ cycles of $T^n$, and the bundle has transition functions in $SL(n;\Z)\subset O(n,n;\Z)$. More generally, one could consider doubled  torus bundles in which the transition functions of the CFT in the fibres involve more general elements of $O(n,n;\Z)$, possibly including strict T-dualities that exchange cycles of $T^n$ and $\widetilde{T}^n$. Such
generalizations include $H$-flux compactifications, where the transition function includes a gauge transformation of the $B$-field (the NS two-form) and T-folds, where the transition functions cannot be understood as a geometric action on the $T^n$ fibres of the torus bundle. In each contractible patch of the base, the T-fold is simply a conventional torus fibration but the patches are glued together by
transition functions which include strict T-dualities, mixing metric and $B$-field degrees of freedom and, as such, the background is globally non-geometric \cite{Hull ``A geometry for non-geometric string backgrounds'',Hellerman:2002ax,Dabholkar Duality Twists}. In particular, one can distinguish between the metric and $B$-field in a contractible region of $M_d$, but the distinction cannot be made globally. There is therefore no global description of such a background in terms of standard Riemannian geometry; however, the doubled formalism gives a geometric description of the transition functions as elements of the mapping class group of the fibre $O(n,n;\Z)\subset SL(2n;\Z)$. Evidence for the existence of such backgrounds has been produced and explicit CFT constructions, describing interpolating orbifold limits of these backgrounds, have been constructed \cite{Flournoy:2005xe,Kawai:2007qd}.

If the base $M_d$ of the torus fibration is a
circle, then the doubled torus bundle ${\cal T}$ is a $T^{2n}$ bundle over $S^1$ and one may consider performing a (non-isometric) generalized T-duality along the base circle \cite{Dabholkar ``Generalised T-duality and non-geometric backgrounds''}. There is evidence that such a non-isometric T-duality is indeed possible and
that the resulting description of the Physics is not even locally geometric. Such locally non-geometric backgrounds are often called
compactifications with `$R$-flux' \cite{Shelton ``Nongeometric flux compactifications'',Shelton:2006fd}. For an $R$-flux background it is not possible to distinguish between metric and $B$-field degrees of freedom even at a point and a description in terms of Riemannian geometry breaks down even locally. The \emph{only} picture we have of such $R$-flux backgrounds is through the doubled formalism. In order to construct a geometric realization of such a background one must introduce a dual coordinate, conjugate to the winding modes around the base circle of ${\cal T}$, and the background may be thought of as a $T^n$ bundle $\widetilde{\cal T}$ with a base circle $\widetilde{S}^1$, parameterised by this `winding coordinate' \cite{Dabholkar ``Generalised T-duality and non-geometric backgrounds'',Hull ``Gauge Symmetry T-Duality and Doubled Geometry''}
\begin{eqnarray}
T^{2n}\hookrightarrow  {\cal T}   &\qquad&T^{2n}\hookrightarrow  \widetilde{\cal T} \nonumber\\
\downarrow  &\qquad\Leftarrow\text{Generalised T-duality}\Rightarrow\qquad&    \,\,\qquad\quad\downarrow\nonumber\\
\,\,S^1 &\qquad  &\qquad\quad\,\widetilde{S}^1\nonumber
\end{eqnarray}
The doubled torus bundles ${\cal T}$ and $\widetilde{\cal T}$ are both $(2n+1)$-dimensional twisted tori, manifolds of the form $\cG'/\G'$, where $\cG'$ is a $(2n+1)$-dimensional group and $\G'\subset\cG'$ is a discrete sub-group such that $\cG'/\G'$ is compact. A more natural doubled geometry with which to describe  such locally non-geometric backgrounds is the $(2n+2)$-dimensional twisted torus $\cX=\cG/\G$ where $\cG$ is a $(2n+2)$-dimensional Lie group with $\G\subset\cG_L$ a discrete subgroup, acting from the left, such that $\cG/\G$ is compact. The embedding of the doubled torus bundles  ${\cal T}$ and $\widetilde{\cal T}$ in this $(2n+2)$-dimensional geometry is highly non-trivial. One might generalise this $(2n+2)$-dimensional construction  and consider backgrounds with a description as a doubled geometry $\cX$ which are not based on torus fibrations and which have no description in terms of $(2n+1)$-dimensional doubled torus bundles ${\cal T}$ and $\widetilde{\cal T}$, but are based on a general even-dimensional Lie group $\cG$.
Progress in developing an approach in which $U$-duality may appear more naturally also been made \cite{Hull ``Flux compactifications of M-theory
on twisted tori'',Cederwall:2007je,Pacheco:2008ps,Hull Generalised geometry for M-theory}.

\subsection{Non-geometric backgrounds and gauged supergravity}

In this paper we shall be particularly interested in studying the lift of massive, gauged supergravities to string theory and investigating the general features of a world-sheet sigma model description of such theories introduced in \cite{New}. Of particular interest are theories in $(D+d)$ space-time dimensions with  a metric, two-form gauge field $\widehat{B}$, scalar field $\widehat{\Phi}$ and
the Lagrangian
\begin{equation}\label{D+d lagrangian}
{\cal L}_{D+d}=e^{-\widehat{\Phi}}\left( \widehat{R}*1-\d\widehat{\Phi} \wedge *\d\widehat{\Phi} -
\frac{1}{2}\widehat{H}\wedge *\widehat{H} \right)+...
\end{equation}
where $\widehat{H}=\d\widehat{B}+...$. The $+...$ in the expressions for ${\cal L}_{D+d}$ and $\widehat{H}$ denote other possible fields such as Ramond-Ramond fields (in Type II theories) or internal gauge fields (in Type I and Heterotic theories) and fermionic fields that are required to ensure supersymmetry. We shall focus attention on the NS-sector given in (\ref{D+d lagrangian}) and not consider other fields explicitly since our primary concern is to study the interrelation between the metric, dilaton and $B$-field on certain string backgrounds.

An example of the type of theory we shall be studying is the lower-dimensional gauged supergravity related to the theory (\ref{D+d lagrangian}) by compactification on a $D$-dimensional manifold of the form ${\cal N}=G/\G$, where $G$ is a $D$-dimensional Lie group and $\G\subset G$ is a discrete sub-group, acting from the left, such that ${\cal N}$ is compact \cite{Hull ``Flux compactifications of string theory on twisted tori'',Hull ``Flux compactifications of M-theory on twisted tori'',ReidEdwards ``Geometric and non-geometric compactifications of IIB supergravity''}. As mentioned above, such geometries are often, if misleadingly, called twisted tori and we shall adopt this name here. The twisted tori are paralleisable and admit a consistent truncation of the field content of the higher-dimensional theory\footnote{See \cite{Hull ``Flux compactifications of string theory on twisted tori'',ReidEdwards ``Geometric and non-geometric compactifications of IIB supergravity'',ReidEdwards:2008rd} for further discussion on this issue.}. Supersymmetry is not explicitly broken by the compactification and the $d$-dimensional effective theory will have sixteen supercharges, if (\ref{D+d lagrangian}) is a Type I or Heterotic theory, or thirty-two supercharges if (\ref{D+d lagrangian}) is a Type II theory; however, supersymmetry may be spontaneously broken for a given vacuum solution of the $d$-dimensional theory. The $(D+d)$-dimensional geometry is then the bundle $Y$, where
\begin{eqnarray}
{\cal N}\hookrightarrow & Y &\nonumber\\
&\downarrow&\nonumber\\
&M_d&\nonumber
\end{eqnarray}
We choose local coordinates on the fibre ${\cal N}$ to be $x^i$, where $i,j=1,2,..D$ and the base $M_d$ has local coordinates $z^{\mu}$, where $\mu,\nu=D+1,...D+d$. The metric and $B$-field reduction ansatze are \cite{Kaloper ``The O(dd) story of massive supergravity''}
\begin{equation}\label{reduction ansatz}
ds_{D+d}=g_{\mu\nu}\d z^{\mu}\otimes \d z^{\nu}+g_{mn}\nu^m\otimes\nu^n \qquad  \widehat{B}=B_{(2)}+B_{(1)m}\wedge\nu^m+\frac{1}{2}B_{mn}\nu^m\wedge\nu^n+\omega
\end{equation}
where $\nu^m$ are related to the left-invariant one-forms $P^m=P^m{}_i\d x^i$ of the group $G$ by
\begin{equation}\label{nu}
\nu^m=P^m+A^m
\end{equation}
where we have introduced the one-forms $A^m=A^m{}_{\mu}\d z^{\mu}$ which have field strength
$$
F^m=\d A^m+\frac{1}{2}f_{np}{}^mA^n\wedge A^p
$$
where $f_{mn}{}^p$ are structure constants for the group $G$. The left-invariant one-forms $P^m$ satisfy the Maurer-Cartan equations
$$
\d P^m+\frac{1}{2}f_{np}{}^mP^n\wedge P^p=0
$$
and are globally defined on both $G$ and ${\cal N}$. We also chose to introduce a flux for the $H$-field such that $\omega$ in (\ref{reduction ansatz}) satisfies
\begin{equation}\label{H-flux}
\d\omega=\frac{1}{6}K_{mnp}P^m\wedge P^n\wedge P^p
\end{equation}
Inserting the reduction ansatz (\ref{reduction ansatz}) and (\ref{H-flux}) into (\ref{D+d lagrangian}) gives the action $S_{D+d}=\int_{Y}{\cal L}_{D+d}=\text{Vol}_{\cal N}\int_{M_d}{\cal L}_d$, where the Lagrangian for the lower-dimensional theory is given by \cite{Kaloper ``The O(dd) story of massive supergravity''}
\begin{eqnarray}\label{D dim sugra}
{\cal L}_d&=&e^{-\phi}\left(R*1+*\d\phi\wedge
\d\phi-\frac{1}{2}H\wedge
*H+\frac{1}{4}*D{\cal M}_{MN}\wedge D{\cal
M}^{MN}\right.
\nonumber\\
&&- \left.\frac{1}{2}{\cal M}_{MN}{\cal
F}^M\wedge*{\cal F}^N\right)-V*1
\end{eqnarray}
The Lagrangian (\ref{D dim sugra}) has a $2D$-dimensional gauge group\footnote{The gauge group is denoted by $\cG_R$ as it may be understood as the right action of the $2D$-dimensional group manifold $\cG$ on itself.} $\cG_R$ and admits a natural rigid action of the group $O(D,D)$. The coset space $O(D,D)/O(D)\times O(D)$ is
parameterised by the symmetric  $2D \times 2D $ matrix of scalar fields  ${\cal M}_{MN}$, which satisfies the constraint
\begin{equation}\label{M=LML2}
{\cal M}_{MN}=L_{MP}({\cal M}^{-1})^{PQ}L_{NQ}
\end{equation}
where $L_{MN}$ is the constant $O(D,D)$-invariant metric, which is used to raise and lower the indices $M,N=1,...,2D$. In terms of the metric and $B$-field on the internal space which appear in the reduction ansatz (\ref{reduction ansatz}), the matrix ${\cal M}_{MN}$ and the one-forms ${\cal A}^M={\cal A}^M{}_{\mu}\d z^{\mu}$ may be written as
$$
{\cal M}_{MN}=\left(
                \begin{array}{cc}
                  g_{mn}-B_{mp}g^{pq}B_{qn} & -g^{np}B_{pm} \\
                  -g^{mp}B_{pn} & g^{mn} \\
                \end{array}
              \right)   \qquad  {\cal A}^M{}_{\mu}=\left(
             \begin{array}{c}
               A^m{}_{\mu} \\
               B_{\mu m} \\
             \end{array}
           \right)
$$
The one-forms ${\cal A}^M$ satisfy
\begin{equation}\label{constraint of A}
{\cal A}^M=L^{MN}{\cal M}_{NP}*{\cal A}^P
\end{equation}
and the scalar potential $V(\phi,{\cal M})$ is given by
$$
V= e^{-\phi}\left(\frac{1}{12}{\cal M}^{MQ}{\cal M}^{NT}{\cal
M}^{PS}t_{MNP}t_{QTS}- \frac{1}{4}{\cal
M}^{MQ}L^{NT}L^{PS}t_{MNP}t_{QTS}\right)
$$
where $t_{MNP}$ are the structure constants of a $2D$-dimensional Lie algebra for the gauge group $\cG$. The kinetic term for the one-forms ${\cal A}^M$ is given in the Lagrangian (\ref{D dim sugra}) in terms of the two-form field strengths
$$
{\cal F}^M=\d{\cal A}^M-\frac{1}{2}t_{NP}{}^M{\cal A}^N\wedge{\cal A}^P
$$
and these one-forms are connections for the gauge group $\cG$. Finally, the three-form field strength $H$ is given by
\begin{equation}\label{C def}
H=\d C_{(2)}-\frac{1}{2}\Omega_{\text{cs}}  \qquad  \text{where}    \qquad
C_{(2)}=B_{(2)}+\frac{1}{2}B_{(1)m}\wedge A^m
\end{equation}
and $\Omega_{\text{cs}}$ is the Chern-Simons term
\begin{equation}\label{chern simons}
\Omega_{\text{cs}}=L_{MN}\left({\cal A}^M\wedge \d{\cal A}^N-\frac{1}{3}t_{PQ}{}^M{\cal A}^N\wedge{\cal A}^P\wedge{\cal A}^Q\right)
\end{equation}
which satisfies $\d\Omega_{\text{cs}}=L_{MN}{\cal F}^M\wedge{\cal F}^N$.

The theory (\ref{D dim sugra}) is invariant under rigid $O(D,D)$ transformations, provided the
 structure constants are taken to transform as a tensor under $O(D,D)$. The generators ${\cal Z}_M$ of the non-abelian gauge symmetry $\cG_R$ consist of $Z_m$, ($m,n=1,2,...D$)  which generate the   right action $G_R$ of $G$ on
the internal space ${\cal N}$, and
 $X^m$, which generate anti-symmetric tensor transformations for the
$B$-field components with one leg on the internal space ${\cal N}$ and the other in the external space-time $M_d$,
 so that
$$
{\cal Z}_M=\left(%
\begin{array}{c}
  Z_m \\
  X^m \\
\end{array}%
\right)
$$
is an $O(D,D)$-vector. The Lie algebra of the gauge symmetry $\cG_R$ can be written as
\begin{equation}\label{doubled algebra}
[\mathcal{Z}_M,\mathcal{Z}_N]=t_{MN}{}^P\mathcal{Z}_P
\end{equation}
and the structure constants satisfy the Jacobi identity $t_{[MN}{}^Qt_{P]Q}{}^T=0$ which encodes the Jacobi identity for $G$; $f_{[mn}{}^qf_{p]q}{}^t=0$ and the identity $K_{[mn|t}f_{|pq]}{}^t=0$ which comes from $\d^2\omega=0$. In this basis, the invariant metric of $O(D,D)$ is off-diagonal
\begin{equation}
\label{Lis}
L_{MN}=\left(\begin{array}{cc}0 & \delta_m{}^n \\ \delta^m{}_n & 0
\end{array}\right)
\end{equation}
The non-vanishing structure constants are $t_{mn}{}^p=-f_{mn}{}^p$, the structure constants for $G$, which encode the local structure of ${\cal N}$ and $t_{mnp}=-K_{mnp}$ which is the constant $H$-flux (\ref{H-flux}). The gauge algebra\footnote{In \cite{Hull ``Flux compactifications of string theory on twisted tori''} it was shown that the true gauge group of (\ref{D dim sugra}) is in fact a generalisation of a Lie group and that the algebra (\ref{algebra}) is in fact the largest Lie sub-algebra of this more general symmetry group. Such subtleties will not play an important role here and will not be considered further.} is
\begin{equation}\label{algebra}
[Z_m,Z_n]=-f_{mn}{}^pZ_p+K_{mnp}X^p    \qquad  [X^m,Z_n]=-f_{np}{}^mX^p    \qquad  [X^m,X^n]=0
\end{equation}
If $K_{mnp}=0$, then the gauge group is simply the semi-direct product $\cG=G\ltimes \R^D$ which is the cotangent bundle $\cG=T^*G$ and the local structure of the internal geometry is recovered from the doubled group by the standard bundle projection $\pi:\cG\rightarrow G$. We shall review the generalisation of this projection for more general $\cG$ in section three.

An important observation is that the Lagrangian (\ref{D dim sugra}) also describes theories which have a more general gauge group, given by a Lie algebra of the form
$$
[Z_m,Z_n]=-f_{mn}{}^pZ_p+K_{mnp}X^p    \qquad  [X^m,Z_n]=-h_{np}{}^mX^p+c_n{}^{mp}Z_p
$$
\begin{equation}\label{general algebra}
[X^m,X^n]=Q^{mn}{}_pX^p+R^{mnp}Z_p
\end{equation}
If we require that (\ref{D+d lagrangian}) describes a sub-sector of some supersymmetric theory then one must consider how the additional fields (e.g Ramond-Ramond fields) transform under the gauge symmetry and the requirement that the gauging preserves supersymmetry places additional constraints on the allowed gauge groups. We will not consider these additional constraints here, but such issues have been studied extensively, using the embedding tensor formalism\footnote{See \cite{Samtleben:2008pe} and references contained therein.}.

An issue of particular importance is whether or not these more general $d$-dimensional gauged theories with Lie algebras of the form (\ref{general algebra}) can be lifted to a compactification of a $(D+d)$-dimensional field theory, such as a supergravity, or string theory. It was argued in \cite{Shelton ``Nongeometric flux compactifications'',Hull ``Flux compactifications of string theory on twisted tori'',Dabholkar Duality Twists} that, whilst many such gauged theories can be lifted to a flux compactification of supergravity theory on a conventional manifold, there are many which cannot. In particular, there are many examples of gauged supergravities for which $Q_m{}^{np}\neq 0$ which do not have a higher dimensional supergravity origin but can only be understood as string theory on a T-fold background \cite{Hull ``Gauge Symmetry T-Duality and Doubled Geometry''}. As reviewed above, such T-fold backgrounds look locally like a conventional Riemannian geometry, but are patched together globally by T-dualities. Furthermore if $R^{mnp}\neq 0$, so that the $X^m$ do not close to form a sub-algebra, then the string background is not even locally geometric. These issues will be reviewed in section three and a more involved discussion may be found in \cite{New}.

In \cite{New,Hull ``Gauge Symmetry T-Duality and Doubled Geometry''} a description of such exotic internal backgrounds was given in terms of a $2D$-dimensional doubled geometry $\cX$ in which the metric and $B$-field degrees of freedom in the conventional $D$-dimensional internal background appear in a T-duality covariant way. This doubled formalism is the natural framework to discuss such non-geometric backgrounds and suggests a generalisation of T-duality beyond the isometric constructions of \cite{Buscher ``A Symmetry of the String Background Field Equations''}. Evidence for such a non-isometric T-duality was given in \cite{Dabholkar ``Generalised T-duality and non-geometric backgrounds''}.

In this paper we shall study a particular example of a six-dimensional doubled twisted torus $\cX$ as a target space for the sigma model introduced in \cite{New} in detail. We shall argue that this sigma model provides a world-sheet description of the $(D+d)$-dimensional string theory lift of examples of massive supergravities of the form (\ref{D dim sugra}). Section two introduces two three-dimensional backgrounds which are related to each other by a fibre-wise T-duality; a nilmanifold with constant $H$-flux and a `twisted' T-fold. In section three the doubled formalism of \cite{Hull ``Gauge Symmetry T-Duality and Doubled Geometry''} will be reviewed and in section four the three-dimensional backgrounds studied in section two will be seen to emerge from two different polarisations of the doubled geometry $\cX$. A third polarisation is also possible and we argue that it gives rise to what might be called a T-fold with $R$-flux. In section five we show how the sigma model describing a closed world-sheet embedding into $\cX$ gives the doubled torus sigma model of \cite{Hull ``A geometry for non-geometric string backgrounds''} as a special case. Finally, in section six, the sigma model, describing the embedding of a world-sheet into $\cX$ is generalised to one describing a target space which is a bundle of $\cX$ over a base $M_d$. Such sigma models form the basis of a world-sheet description of the field theory $(\ref{D dim sugra})$.

\section{Nilmanifold with $H$-flux and its duals}

Consider the background given by a particular three-dimensional nilmanifold with a constant $H$-flux
\begin{equation}\label{nil background}
ds^2=\d x^2+\d y^2+(\d z+nx\d y)^2  \qquad  H=m\d x\wedge \d y\wedge \d z
\end{equation}
where $m,n\in\Z$. Compactification of the field theory (\ref{D+d lagrangian}) on this three-dimensional nilmanifold with a constant $H$-flux gives the non-abelian gauge theory (\ref{D dim sugra}) characterised by the non-vanishing structure constants
$$
t_{xyz}=-m  \qquad  t_{xy}{}^z=-n
$$
The nilmanifold is locally a group, $G$, but globally is a twisted torus of the kind discussed in the previous section and takes the form ${\cal N}=G/\G$, where $\G\subset G_L$ is a discrete (cocompact) subgroup of $G$ which acts from the left. The nilmanifold may be written in terms of left-invariant one-forms on $G$ as
$$
\ell^x=\d x  \qquad  \ell^y=\d y  \qquad  \ell^z=\d z+nx\d y
$$
The local coordinates are identified under $\G$ as
$$
(x,y,z)\sim (x+1,y,z-ny)  \qquad  (x,y,z)\sim (x,y+1,z)   \qquad  (x,y,z)\sim (x,y,z+1)
$$
and we see that the left-invariant one-forms $\ell^m$ are well-defined on $\cal N$. Dual to these one-forms are the left-invariant vector fields
$$
K_x=\frac{\partial}{\partial x} \qquad  K_y=\frac{\partial}{\partial y}-nx\frac{\partial}{\partial z}   \qquad  K_z=\frac{\partial}{\partial z}
$$
which generate $G_R$, the right action of $G$. These vector fields are globally defined on ${\cal N}$, but the explicit $x$-dependence in the metric (\ref{nil background}) means that only $Z_y$ and $Z_z$ are Killing vector fields. Note that $G_R$ is a subgroup of a contraction of the full non-abelian gauge group $\cG$
$$
[Z_x,Z_y]=nZ_z-mX^z \qquad  [Z_y,Z_z]=mX^x  \qquad  [Z_x,Z_z]=-mX^y
$$
$$
[Z_x,X^z]=nX^y  \qquad  [Z_y,X^z]=-nX^x
$$
where all other commutators vanish. If we define $X^m=\lambda W^m$ and take the limit $\lambda\rightarrow 0$ the non-trivial commutators of the algebra are
$$
[Z_x,Z_y]=nZ_z \qquad  [Z_x,Y^z]=nW^y  \qquad  [Z_y,Y^z]=-nW^x
$$
so that the $Z_m$'s close to give a sub-algebra of this contracted algebra.

We can also consider the right-invariant one-forms
$$
r^x=\d x  \qquad  r^y=\d y  \qquad  r^z=\d y+ny\d x
$$
and the dual, right-invariant, vector fields
$$
\widetilde{K}_x=\frac{\partial}{\partial x}-ny\frac{\partial}{\partial z} \qquad  \widetilde{K}_y=\frac{\partial}{\partial y}   \qquad  \widetilde{K}_z=\frac{\partial}{\partial z}
$$
which generate $G_L$, the left action of $G$. The right-invariant forms are well-defined on the group manifold $G$, but generally will not be well-defined on the twisted torus ${\cal N}$. Under the action of $\G$, the right-invariant vector fields transform as
\begin{equation}\label{K tranformations}
\widetilde{K}_x\rightarrow\widetilde{K}_x-n\beta\widetilde{K}_z    \qquad  \widetilde{K}_y\rightarrow\widetilde{K}_y-n\alpha\widetilde{K}_z \qquad  \widetilde{K}_z\rightarrow\widetilde{K}_z
\end{equation}
where $\alpha$ and $\beta$ are integers which parameterise $\G$, so that only ($K_x,K_y,K_z=\widetilde{K}_z$) are globally defined and of these only ($K_y,K_z=\widetilde{K}_z$) are Killing.

A sigma model, describing the embedding of the world-sheet $\Sigma$ into this background, is given by
$$
S_{\cal N}=\frac{1}{2}\oint_{\Sigma}g_{ij}\d x^i\wedge \d x^j +\int_Vmdx\wedge \d y\wedge \d z
$$
where $V$ is a formal three-dimensional extension of the world-sheet such that $\partial V=\Sigma$ and $g_{ij}$ is the left-invariant metric in (\ref{nil background}). $G_L$ is a rigid symmetry of this sigma model.

\subsection{T-duality and T-folds}

The Buscher construction \cite{Buscher ``A Symmetry of the String Background Field Equations''} gives a procedure to find a T-dual description of this nilmanifold with $H$-flux background provided that there exists an isometry which preserves the dilaton\footnote{Throughout this paper we shall assume that the dilaton does not depend on the coordinates $x^i$ of the internal space.} and $H$-field strength\footnote{See \cite{Hull Global Aspects of T-Duality} for a generalisation of the Buscher construction.}. From (\ref{K tranformations}) we see that the only globally-defined generator of $G_L$ is $\widetilde{K}_z$ and it is not hard to show that $\widetilde{K}_z$ does indeed preserve the $H$-field strength:
$$
{\cal L}_z H=(\iota_z\d+\d\iota_z)H=0
$$
where $\iota_z$ denotes contraction with $\widetilde{K}_z$, and we see that it is possible to perform the T-duality along the $z$-direction according to the Buscher prescription. The duality has the effect of exchanging $m$ and $n$ so that the resulting T-dual model is also a nilmanifold with constant $H$-flux, but where
$$
\d s^2=\d x^2+\d y^2+(\d z+mx\d y)^2  \qquad  H=n\d x\wedge \d y\wedge \d z
$$
so that if $m=n$, the model is self-dual when the radius of the $z$-direction is at the self-dual point. If we consider instead a cover of the nilmanifold, given by dropping the identification in the $x$-coordinate $x\sim x+1$, then we find that the generator $\widetilde{K}_y$ is well-defined on this cover and we may then consider performing a T-duality along the $y$-direction. The $H$-field strength is preserved by this vector field $\widetilde{K}_y$, i.e. ${\cal L}_y H=0$ and the dual background is the smooth geometry given by
$$
\d s^2=\d x^2+\frac{1}{1+(nx)^2}(\d y-mx\d z)^2+\frac{1}{1+(nx)^2}\d z^2  \qquad  B=-\frac{nx}{1+(nx)^2}\d y\wedge \d z
$$
We now consider the background given by imposing the identification $x\sim x+1$. If $m=0$, then the background is the familiar three-dimensional T-fold \cite{Hull ``A geometry for non-geometric string backgrounds''}. Conversely, if $n=0$ the background is a nilmanifold, given by a $T^2$ bundle over $S^1_x$. For $m$ and $n$ both non-zero the background is a more general class of T-fold and we might call this background a `twisted T-fold'. We shall consider this background again from the doubled perspective in section three. It is interesting to note that dualising along the $z$-direction simply has the effect of exchanging the roles of $m$ and $n$ as seen above.

Defining the background tensor $E=g+B$, the Buscher rules may be expressed simply as a set of $\Z_2$ subgroups of an $O(3,3;\Z)$ transformation, acting as a fractional-linear transformation
$$
E\rightarrow (M\cdot E +N)^{-1}(J \cdot E+K)    \qquad  \left(
                                                          \begin{array}{cc}
                                                            J & K \\
                                                            M & N \\
                                                          \end{array}
                                                        \right)\in O(3,3;\Z)
$$
where $\cdot$ denotes matrix multiplication. A generic $O(3,3;\Z)$ matrix may be written as a product of the matrices
$$
{\cal O}_A=\left(
             \begin{array}{cc}
               A & 0 \\
               0 & (A^{-1})^T \\
             \end{array}
           \right)  \qquad  {\cal O}_b=\left(
             \begin{array}{cc}
               1 & b \\
               0 & 1 \\
             \end{array}
           \right)  \qquad {\cal O}_{\beta}=\left(
             \begin{array}{cc}
               1 & 0 \\
               \beta & 1 \\
             \end{array}
           \right)
$$
where $A\in GL(3;\Z)$, and $b$ and $\beta$ are antisymmetric $3\times 3$ matrices with integer components. An alternative parameterisation of the $O(3,3;\Z)$ would be to use the set of matrices $({\cal O}_A, {\cal O}_b,{\cal O}_m)$ where
$$
{\cal O}_m=\left(
             \begin{array}{cc}
               1-e_m & e_m \\
               e_m & 1-e_m \\
             \end{array}
           \right)
$$
instead of the set $({\cal O}_A, {\cal O}_b,{\cal O}_{\beta})$ above, where $e_m$ is the diagonal $3\times 3$ matrix with zero along the diagonal, except for a 1 in the $m$'th position. The action of ${\cal O}_m$ is equivalent to applying the Buscher rules using $\widetilde{K}_m$, where $\widetilde{K}_m$ is Killing and preserves the $H$-field. For the three-dimensional backgrounds considered here there is no isometry, even locally, in the $x$-direction, so we cannot use the Buscher rules, even on a cover of the space; however, in section three we shall perform a non-isometric or `generalised duality' along the lines of that of \cite{Dabholkar ``Generalised T-duality and non-geometric backgrounds''} and suggest another dual description of this background. This background, related to the nilmanifold and T-fold described above by the conjectured non-isometric T-duality, will be an example of the $R$-flux backgrounds discussed in the Introduction.

It was argued in \cite{New} that an equivalent description of backgrounds such as the nilmanifold and T-fold backgrounds described above may be given in terms of a six-dimensional doubled twisted torus
$$
\cX=\cG/\G
$$
where $\cG$ is the Lie group with Lie algebra (\ref{doubled algebra}) and $\G\subset\cG_L$ is a discrete (cocompact) subgroup, which acts from the left, such that $\cG/\G$ is compact. The algebra (\ref{doubled algebra}) fixes the local structure of $\cX$ and it is useful to introduce left-invariant one-forms ${\cal P}^M$, dual to the generators ${\cal Z}_M$, which satisfy the Maurer-Cartan structure equations
\begin{equation}\label{Maurer-Cartan}
\d{\cal P}^M+\frac{1}{2}t_{NP}{}^M{\cal P}^N\wedge{\cal P}^P=0
\end{equation}
A conventional, $D$-dimensional, description of the background is recovered from the doubled geometry by choosing a polarisation $\Pi$ which selects a set of $D$ generators $\widetilde{X}^m=\Pi^{mM}\widetilde{{\cal Z}}_M$. Once a polarisation has been chosen, the types of string theory backgrounds that are described by the doubled geometry fall into three categories \cite{New}:

\noindent\underline{Type I: Riemannian geometry}

If the generators $\widetilde{X}^m$ selected by the polarisation generate a (maximally isotropic) subgroup $\widetilde{G}_L\subset\cG_L$ and this sub-group preserves and is preserved by $\G$, then the quotient $\cX/\widetilde{G}_L$ is well-defined. Furthermore, this quotient $\cX/\widetilde{G}_L$ provides a global description of the $D$-dimensional compactification geometry.

\noindent\underline{Type II: T-fold}

If the generators $\widetilde{X}^m$ generate a sub-group $\widetilde{G}_L$ but this sub-group is not preserved by $\G$, then the quotient $\cX/\widetilde{G}_L$ is not well-defined and a global description of the background in terms of a $B$-field on a $D$-dimensional Riemannian geometry is not possible. Locally, $\cX$ looks like $\cG$ and the coset $\cG/\widetilde{G}_L$ is well-defined. Therefore, locally the background is a conventional $D$-dimensional Riemannian geometry and can be recovered locally as a patch of the coset $\cG/\widetilde{G}_L$. A global string theory description of the background  is given by gluing the local string theory descriptions in each geometric patch together by the action of $\G$ on the string theory in the individual patches. In this case, the transition functions will include non-geometric transformations such as strict T-dualities. If the action of $\G$ is a symmetry of the string theory (such as a strict T-duality), then the background is a T-fold and, though not a Riemannian geometry, is a candidate for a smooth string theory background.

\noindent\underline{Type III: Locally non-geometric background}

If the polarisation selects generators $\widetilde{X}^m$ which do not close to form a sub-algebra, then we cannot recover a conventional description of the background even locally. Examples include backgrounds with `$R$-flux'.

We shall return to these issues in section three. A full discussion with further examples can be found in \cite{New}. As described there (see also \cite{Hull ``Gauge Symmetry T-Duality and Doubled Geometry'',Dall'Agata:2007sr}), different polarisations are related by the action of $O(D,D;\Z)$ and give rise to different $D$-dimensional backgrounds. In general, $O(D,D;\Z)$ maps one background to another, physically inequivalent, background; however, if the action of the $O(D,D;\Z)$ is a symmetry of the theory (for example T-duality of tori, or the cases studied in this section), then the different polarisations relate different descriptions of the same physics.

The doubled geometry $\cX$ has a natural metric and three-form given by
$$
\d s^2={\cal M}_{MN}{\cal P}^M\otimes {\cal P}^N  \qquad {\cal K}=\frac{1}{6}t_{MNP}{\cal P}^M\wedge{\cal P}^N\wedge{\cal P}^P
$$
and the action of $O(D,D;\Z)$ has a particularly particularly simple form on the doubled fields
$$
{\cal M}_{MN}\rightarrow {\cal O}_M{}^P{\cal M}_{PQ}{\cal O}^Q{}_N   \qquad  {\cal P}^M\rightarrow ({\cal O}^{-1})^M{}_N{\cal P}^N  \qquad  t_{MNP}\rightarrow t_{QTS}{\cal O}^Q{}_M{\cal O}^T{}_N{\cal O}^S{}_P
$$
where ${\cal O}\in O(D,D;\Z)$. This linear action on the fields on $\cX$ is much simpler that the fractional-linear action of $O(D,D;\Z)$ on the fields $E=g+B$ and suggests a generalisation of the Buscher rules to the non-isometric case \cite{New,Dabholkar ``Generalised T-duality and non-geometric backgrounds'',Hull ``Gauge Symmetry T-Duality and Doubled Geometry''}.

\section{Doubled geometry}

In this section we describe the backgrounds in section two from the perspective of the doubled twisted torus $\cX=\cG/\G$. The six-dimensional doubled group $\cG$ has (matrix) generators $T_M$ which satisfy the Lie algebra
\begin{equation}\label{lie algebra}
[T_M,T_N]=t_{MN}{}^PT_P
\end{equation}
The indices $M,N$ are lowered (raised) by $L_{MN}$ ($L^{MN}$), the invariant of $O(D,D;\Z)$, given by (\ref{Lis}) and $t_{MNP}=L_{MQ}t_{MN}{}^Q$ is totally antisymmetric. Note that $L$ exchanges pairs of indices, so that the non-vanishing structure constants for $\cG$, which may be written as
\begin{equation}\label{structure constants}
t_{123}=-m\in\Z  \qquad  t_{126}=-n\in\Z
\end{equation}
encode the six non-trivial structure constants
\begin{equation}\label{structure constants2}
t_{12}{}^3=t_{26}{}^4=t_{61}{}^5=-n \qquad  t_{23}{}^4=t_{31}{}^5=t_{12}{}^6=-m
\end{equation}
The Lie algebra for $\cG$ may be equivalently written as the Maurer-Cartan (\ref{Maurer-Cartan}) equations for a basis of left-invariant one-forms ${\cal P}={\cal P}^MT_M=g^{-1}\d g$, where $g\in\cG$ and, for the structure constants (\ref{structure constants2}), are
\begin{eqnarray}\label{6 dim MC}
\begin{array}{lll}
\d{\cal P}^1=0  &\qquad \d{\cal P}^3-n{\cal P}^1\wedge {\cal P}^2=0 &\qquad    \d{\cal P}^5+n{\cal P}^1\wedge{\cal P}^6-m{\cal P}^3\wedge{\cal P}^1=0\\
 \d{\cal P}^2=0  &\qquad \d{\cal P}^4-n{\cal P}^2\wedge{\cal P}^6+m{\cal P}^3\wedge{\cal P}^2=0    &\qquad \d{\cal P}^6+m{\cal P}^2\wedge{\cal P}^1=0
\end{array}
\end{eqnarray}
We can define local coordinates $\mathbb{X}^I$ ($I=1,2,...6$) on $\cG$ and solve these Maurer-Cartan equations to give local descriptions of the left-invariant one-forms ${\cal P}^M={\cal P}^M{}_I\d\XX^I$ in terms of these coordinates
\begin{eqnarray}\label{left invariant forms}
\begin{array}{ll}
{\cal P}^1=\d\mathbb{X}^1  &\qquad  {\cal P}^4=\d\mathbb{X}^4+m\mathbb{X}^2\d\XX^3+n\mathbb{X}^2\d\XX^6+mn\XX^1\mathbb{X}^2\d\XX^2  \\
{\cal P}^2=\d\mathbb{X}^2  &\qquad  {\cal P}^5=\d\mathbb{X}^5-m\mathbb{X}^1\d\XX^3-n\mathbb{X}^1\d\XX^6+mn\mathbb{X}^1\XX^2\d\XX^1   \\
{\cal P}^3=\d\mathbb{X}^3+n\mathbb{X}^1\d\XX^2 &\qquad {\cal P}^6=\d\mathbb{X}^6-m\mathbb{X}^2\d\XX^1
\end{array}
\end{eqnarray}
Using this basis we may then find explicit expressions for the left-invariant vector fields ${\cal Z}_M=({\cal P}^{-1})_M{}^I\partial_I$, which generate $\cG_R$, the right action of the doubled group and satisfy the Lie algebra (\ref{doubled algebra})
$$
{\cal Z}_1=\frac{\partial}{\partial\mathbb{X}^1} - mn(\mathbb{X}^2)^2\frac{\partial}{\partial\mathbb{X}^4}+m\mathbb{X}^2\frac{\partial}{\partial\mathbb{X}^5} \quad,   \quad  {\cal Z}_2=\frac{\partial}{\partial\mathbb{X}^2} - mn(\mathbb{X}^1)^2\frac{\partial}{\partial\mathbb{X}^5}-n\mathbb{X}^1\frac{\partial}{\partial\mathbb{X}^3}
$$
$$
{\cal Z}_3=\frac{\partial}{\partial\mathbb{X}^3} - m\mathbb{X}^2\frac{\partial}{\partial\mathbb{X}^4}+m\mathbb{X}^1\frac{\partial}{\partial\mathbb{X}^5}
$$
$$
{\cal Z}_4=\frac{\partial}{\partial\mathbb{X}^4}   \quad,   \quad  {\cal Z}_5=\frac{\partial}{\partial\mathbb{X}^5} \quad,   \quad {\cal Z}_6=\frac{\partial}{\partial\mathbb{X}^6}-m\mathbb{X}^2\frac{\partial}{\partial\mathbb{X}^4}+n\mathbb{X}^1\frac{\partial}{\partial\mathbb{X}^5}
$$
These vector fields are dual to the left-invariant one-forms
$$
\langle{\cal Z}_M,{\cal P}^N\rangle=\delta_M{}^N
$$
where $\langle \partial_I,\d\mathbb{X}^J\rangle=\delta_I{}^J$ is the natural, $O(3,3;\Z)$ and adjoint-invariant inner product between forms and vectors. Both the left-invariant one-forms and vector fields are invariant under the action of $\cG_L$.

One may also define a basis of right-invariant one-forms $\widetilde{\cal P}=\widetilde{\cal P}^MT_M=\d gg^{-1}$, which satisfy the Maurer-Cartan equations
\begin{eqnarray}
\begin{array}{lll}
\d\widetilde{{\cal P}}^1=0  &\qquad \d\widetilde{{\cal P}}^3+n\widetilde{{\cal P}}^1\wedge \widetilde{{\cal P}}^2=0 &\qquad    \d\widetilde{{\cal P}}^5-n\widetilde{{\cal P}}^1\wedge\widetilde{{\cal P}}^6+m\widetilde{{\cal P}}^3\wedge\widetilde{{\cal P}}^1=0\\
 \d\widetilde{{\cal P}}^2=0  &\qquad \d\widetilde{{\cal P}}^4+n\widetilde{{\cal P}}^2\wedge\widetilde{{\cal P}}^6-m\widetilde{{\cal P}}^3\wedge\widetilde{{\cal P}}^2=0    &\qquad \d\widetilde{{\cal P}}^6-m\widetilde{{\cal P}}^2\wedge\widetilde{{\cal P}}^1=0
\end{array}\nonumber
\end{eqnarray}
which are identical in form to those in (\ref{6 dim MC}), except that $(m,n)$ is replaced by $(-m,-n)$. These Maurer-Cartan equations may be solved and written in terms of the same set of local coordinates as
\begin{eqnarray}\label{right inv doubled forms}
\begin{array}{ll}
\widetilde{{\cal P}}^1=\d\mathbb{X}^1  &\qquad  \widetilde{{\cal P}}^4=\d\mathbb{X}^4+m\mathbb{X}^3\d\mathbb{X}^2+n\mathbb{X}^6\d\mathbb{X}^2+mn\mathbb{X}^1\mathbb{X}^2\d\mathbb{X}^2  \\
\widetilde{{\cal P}}^2=\d\mathbb{X}^2  &\qquad  \widetilde{{\cal P}}^5=\d\mathbb{X}^5-m\mathbb{X}^3\d\mathbb{X}^1-n\mathbb{X}^6\d\mathbb{X}^1+mn\mathbb{X}^1\mathbb{X}^2\d\mathbb{X}^1   \\
\widetilde{{\cal P}}^3=\d\mathbb{X}^3+n\mathbb{X}^2\d\mathbb{X}^1 &\qquad \widetilde{{\cal P}}^6=\d\mathbb{X}^6-m\mathbb{X}^1\d\mathbb{X}^2
\end{array}
\end{eqnarray}
The right-invariant one-forms $\widetilde{\cal P}^M=\widetilde{\cal P}^M{}_I\d\mathbb{X}^I$ are dual to the right-invariant vector fields $\widetilde{\cal Z}_M=(\widetilde{\cal P}^{-1})_M{}^I\partial_I$ which generate $\cG_L$, the left action of $\cG$, where $\langle\widetilde{\cal Z}_M,\widetilde{\cal P}^N\rangle=\delta_M{}^N$. In the coordinates given above, these vector fields may be written as
$$
\widetilde{\cal Z}_1=\frac{\partial}{\partial\mathbb{X}^1} -n\mathbb{X}^2\frac{\partial}{\partial\mathbb{X}^3} +(m\mathbb{X}^3+n\mathbb{X}^6-mn\mathbb{X}^1\mathbb{X}^2)\frac{\partial}{\partial \mathbb{X}^5}
$$
$$
\widetilde{\cal Z}_2=\frac{\partial}{\partial\mathbb{X}^2} +m\mathbb{X}^1\frac{\partial}{\partial\mathbb{X}^6} -(m\mathbb{X}^3+n\mathbb{X}^6+mn\mathbb{X}^1\mathbb{X}^2)\frac{\partial}{\partial \mathbb{X}^4}
$$
$$
{\cal Z}_3=\frac{\partial}{\partial\mathbb{X}^3}  \quad,   \quad {\cal Z}_4=\frac{\partial}{\partial\mathbb{X}^4}
$$
$$
{\cal Z}_5=\frac{\partial}{\partial\mathbb{X}^5} \quad,   \quad {\cal Z}_6=\frac{\partial}{\partial\mathbb{X}^6}
$$
and satisfy the commutation relations
$$
[\widetilde{\mathcal{Z}}_M,\widetilde{\mathcal{Z}}_N]=-t_{MN}{}^P\widetilde{\mathcal{Z}}_P
$$

\subsection{Doubled twisted torus}

The geometry we are interested in is not the non-compact group $\cG$, but the twisted torus $\cX=\cG/\G$, where $\G\subset\cG_L$ is a discrete sub-group of $\cG$ which acts from the left such that $\cX$ is compact. Such discrete groups are said to be cocompact. We can define the group $\G$ by its action on the coordinates of the group $\cG$ given above. The identification
$$
g\sim hg    \qquad  g\in\cG, h\in\G
$$
imposes the identifications on the coordinates
\begin{eqnarray}
(\mathbb{X}^1,\mathbb{X}^2,\mathbb{X}^3,\mathbb{X}^4,\mathbb{X}^5,\mathbb{X}^6)&\sim&(\mathbb{X}^1+1,\mathbb{X}^2,\mathbb{X}^3-n\mathbb{X}^2,\mathbb{X}^4,
\mathbb{X}^5+(m\mathbb{X}^3+n\mathbb{X}^6-mn\mathbb{X}^1\mathbb{X}^2)-mn\mathbb{X}^2,\mathbb{X}^6)\nonumber\\
(\mathbb{X}^1,\mathbb{X}^2,\mathbb{X}^3,\mathbb{X}^4,\mathbb{X}^5,\mathbb{X}^6)
&\sim&(\mathbb{X}^1,\mathbb{X}^2+1,\mathbb{X}^3,\mathbb{X}^4-(m\mathbb{X}^3+n\mathbb{X}^6+mn\mathbb{X}^1\mathbb{X}^2)-mn\mathbb{X}^1 ,
\mathbb{X}^5,\mathbb{X}^6+m\mathbb{X}^1)\nonumber\\
(\mathbb{X}^1,\mathbb{X}^2,\mathbb{X}^3,\mathbb{X}^4,\mathbb{X}^5,\mathbb{X}^6)&\sim&(\mathbb{X}^1,\mathbb{X}^2,\mathbb{X}^3+1,\mathbb{X}^4,\mathbb{X}^5,\mathbb{X}^6)\nonumber\\
(\mathbb{X}^1,\mathbb{X}^2,\mathbb{X}^3,\mathbb{X}^4,\mathbb{X}^5,\mathbb{X}^6)&\sim&(\mathbb{X}^1,\mathbb{X}^2,\mathbb{X}^3,\mathbb{X}^4+1,\mathbb{X}^5,\mathbb{X}^6)\nonumber\\
(\mathbb{X}^1,\mathbb{X}^2,\mathbb{X}^3,\mathbb{X}^4,\mathbb{X}^5,\mathbb{X}^6)&\sim&(\mathbb{X}^1,\mathbb{X}^2,\mathbb{X}^3,\mathbb{X}^4,\mathbb{X}^5+1,\mathbb{X}^6)\nonumber\\
(\mathbb{X}^1,\mathbb{X}^2,\mathbb{X}^3,\mathbb{X}^4,\mathbb{X}^5,\mathbb{X}^6)&\sim&(\mathbb{X}^1,\mathbb{X}^2,\mathbb{X}^3,\mathbb{X}^4,\mathbb{X}^5,\mathbb{X}^6+1)\nonumber
\end{eqnarray}
The left- and right-invariant one-forms and vector fields are globally well-defined on $\cG$. The left-invariant one-forms ${\cal P}^M$ and vector fields ${\cal Z}_M$ are invariant under the action of $\G\subset\cG_L$ and are therefore remain well-defined on the twisted torus $\cX$; however, the right-invariant one-forms $\widetilde{\cal P}^M$ and vector fields $\widetilde{\cal Z}_M$ will in general not be preserved by the action of $\G$ and will therefore not be well-defined on $\cX$. In fact one can show that, under the action of $\G$ as defined on the coordinates above, the right-invariant vector fields change as
\begin{eqnarray}\label{generator transformations}
\widetilde{{\cal Z}}_1&\rightarrow& \widetilde{{\cal Z}}_1-n\beta\widetilde{{\cal Z}}_3-m\beta\widetilde{{\cal Z}}_6 -mn\beta\widetilde{{\cal Z}}_4 +(m\gamma+n\tilde{\gamma}+mn\alpha\beta)\widetilde{{\cal Z}}_5\nonumber\\
\widetilde{{\cal Z}}_2&\rightarrow& \widetilde{{\cal Z}}_2-n\alpha\widetilde{{\cal Z}}_3-m\alpha\widetilde{{\cal Z}}_6 -mn\alpha^2\widetilde{{\cal Z}}_5 +(m\gamma+n\tilde{\gamma}-mn\alpha\beta)\widetilde{{\cal Z}}_4\nonumber\\
\widetilde{{\cal Z}}_3&\rightarrow &\widetilde{{\cal Z}}_3+m\beta\widetilde{{\cal Z}}_4-m\alpha\widetilde{{\cal Z}}_5 \nonumber\\
\widetilde{{\cal Z}}_4&\rightarrow &\widetilde{{\cal Z}}_4\nonumber\\
\widetilde{{\cal Z}}_5&\rightarrow &\widetilde{{\cal Z}}_5\nonumber\\
\widetilde{{\cal Z}}_6&\rightarrow& \widetilde{{\cal Z}}_6+n\beta\widetilde{{\cal Z}}_4-n\alpha\widetilde{\cal Z}_5
\end{eqnarray}
where $(\alpha,\beta,\gamma,\tilde{\alpha},\tilde{\beta},\tilde{\gamma})$ are constant integers which parameterise the discrete group $\G$. We see that the only maximally isotropic sub-groups, which are preserved by this identification are those generated by:
$$
(\widetilde{{\cal Z}}_4,\widetilde{{\cal Z}}_5,\widetilde{{\cal Z}}_6)\qquad \text{and}  \qquad  (\widetilde{{\cal Z}}_3,\widetilde{{\cal Z}}_4,\widetilde{{\cal Z}}_5)
$$
The existence of maximally isotropic sub-groups which are preserved by $\G$ is of great importance in recovering globally geometric backgrounds in various polarisations. A polarisation that selects either of these two sub-groups will be associated to a conventional Riemannian geometry or, using the categories described in section 2.1, a Type I background. The fact that there are only two such sub-groups indicates that there are only two polarisations which will give rise to a globally geometric space-time and it not hard to see that the two sets of globally-defined generators are exchanged by a T-duality along the $z$-direction which simply exchanges $m$ and $n$, which in turn exchanges $\widetilde{\cal Z}_3$ with $\widetilde{{\cal Z}}_6$, and therefore interchanges the two sets of generators $
(\widetilde{{\cal Z}}_4,\widetilde{{\cal Z}}_5,\widetilde{{\cal Z}}_6)$ and $(\widetilde{{\cal Z}}_3,\widetilde{{\cal Z}}_4,\widetilde{{\cal Z}}_5)
$.

\subsubsection{Polarisations and the existence of space-time}

The conventional description of the background is recovered by choosing a polarisation, as mentioned briefly in section 2.1 and discussed in more detail in \cite{New}. See \cite{Hull ``A geometry for non-geometric string backgrounds''} for a discussion of polarisations relating to the doubled torus construction. The polarisation, $\Pi$, selects a set of generators $X^m$ and one-forms $P^m$
$$
P^m=\Pi^m{}_M{\cal P}^M \qquad  X^m=\Pi^{mM}{\cal Z}_M \qquad \text{where}    \qquad \Pi^{mM}=\Pi^m{}_NL^{NM}
$$
And similarly for the right-invariant fields
$$
\widetilde{P}^m=\Pi^m{}_M\widetilde{{\cal P}}^M \qquad  \widetilde{X}^m=\Pi^{mM}\widetilde{{\cal Z}}_M
$$
If the generators close to form a subalgebra $[\widetilde{X}^m,\widetilde{X}^n]=f^{mn}{}_p\widetilde{X}^p$ then, by Frobenius' theorem, the polarisation defines a three-dimensional sub-manifold $\widetilde{G}\subset\cG$ \cite{New}. The integrability condition for the existence of $\widetilde{G}$ is then
$$
\Pi^{mM}\Pi^{nN}\Pi^p{}_Pt_{MN}{}^P:=R^{mnp}=0
$$
It is useful (although not necessary) to define a complementary polarisation $\widetilde{\Pi}$ such that $Q_m=\widetilde{\Pi}_{mM}{\cal P}^M$, $Z_m=\widetilde{\Pi}_m{}^M{\cal Z}_M$ and similarly for the right-invariant fields. These polarisations may be combined into a polarisation tensor
$$
 \Theta^{\hat M}{}_M=\left(
                        \begin{array}{c}
                          \Pi^m{}_M \\
                          \widetilde{\Pi}_{mM} \\
                        \end{array}
                      \right)   \qquad  \text{so that}  \qquad   {\cal P}^{\hat M}:=\Theta^{\hat M}{}_M{\cal P}^M=\left(
                        \begin{array}{c}
                          P^m \\
                          Q_m \\
                        \end{array}
                      \right)
$$
This in turn defines a local polarisation for the coordinates
$$
x^i=\Pi^i{}_I\mathbb{X}^I    \qquad  {\ti x}_i=\widetilde{\Pi}_{iI}\mathbb{X}^I
$$
where we take ${\ti x}_i=(\tilde{x},\tilde{y},\tilde{z})$ to be local coordinates on the group $\widetilde{G}$ and $x^i=(x,y,z)$ are the coordinates on the coset $\cG/\widetilde{G}$. A patch of this coset gives a local description of the three-dimensional internal geometry.

\subsection{Non-linear sigma-model for the doubled twisted torus}

\noindent The action describing the embedding of a closed string world-sheet $\Sigma$ into the target space $\cX$ is \cite{New}
\begin{eqnarray}\label{fibre action}
S_{\cX}&=&\frac{1}{4}\oint_{\Sigma}\d^2\sigma\sqrt{h}h^{\alpha\beta}{\cal H}_{IJ}\partial_{\alpha}\mathbb{X}^I\partial_{\beta}\mathbb{X}^J +\frac{1}{12}\int_V\d^3\sigma'\varepsilon^{\alpha'\beta'\gamma'}{\cal K}_{IJK}\partial_{\alpha'}\mathbb{X}^I\partial_{\beta'}\mathbb{X}^J\partial_{\gamma'}\mathbb{X}^K\nonumber\\
&&+\frac{1}{2\pi}\oint_{\Sigma}\d^2\sigma\sqrt{h}\phi R(h)
\end{eqnarray}
where $V$ is an extension of the world-sheet, with coordinates $\sigma^{\alpha'}$, such that $\partial V=\Sigma$. We shall choose a gauge in which the world-sheet metric $h_{\alpha\beta}$ is flat and Lorentzian and so the world-sheet Ricci scalar
$R(h)$ is zero and the world-sheet Hodge star is an almost product structure $*^2=+1$. The target space metric ${\cal H}_{IJ}={\cal H}_{IJ}(\mathbb{X})$ and Wess-Zumino
three-form are given by
$$
{\cal H}_{IJ}={\cal M}_{MN}{\cal P}^M{}_I{\cal P}^N{}_J    \qquad  {\cal K}_{IJK}=t_{MNP}{\cal P}^M{}_I{\cal P}^N{}_J{\cal P}^P{}_K
$$
so that the line element and three-form on the twisted torus $\cX$ may be written as
$$
\d s_{\cX}^2={\cal M}_{MN}{\cal P}^M\otimes {\cal P}^N    \qquad  {\cal K}=\frac{1}{6}t_{MNP}{\cal P}^M\wedge {\cal P}^N\wedge{\cal
P}^P
$$
where ${\cal P}=g^{-1}\d g$ is the pull-back of the left-invariant one-forms (\ref{left invariant forms}) to $\Sigma$, where now $\d=\d\sigma^{\alpha}\partial_{\alpha}$ is a world-sheet derivative and the one-forms satisfy the world-sheet Maurer-Cartan equations
\begin{equation}\label{worldsheet MC equations}
\d{\cal P}^M+\frac{1}{2}t_{NP}{}^M{\cal P}^N\wedge{\cal P}^P=0
\end{equation}
$\mathcal{M}_{MN}$ is the matrix given in (\ref{D dim sugra}) that parameterises the coset $O(D,D)/O(D)\times
O(D)$ and is independent of $\mathbb{X}^I$ and $t_{MNP}=L_{MQ}t_{NP}{}^Q$ are the structure constants for the Lie algebra (\ref{doubled algebra}). We
can write the Wess-Zumino field strength as ${\cal K}_{IJK}=t_{MNP}{\cal P}^M{}_I{\cal P}^N{}_J{\cal P}^P{}_K=t_{MNP}\widetilde{{\cal
P}}^M{}_I\widetilde{{\cal P}}^N{}_J\widetilde{{\cal P}}^P{}_K$, where $\widetilde{\cal P}=\d gg^{-1}$ is the pull-back of the right-invariant one-forms (\ref{right inv doubled forms}) to $\Sigma$. We see then that the sigma model has a manifest, left-acting $\cG_L$ symmetry.  The Wess-Zumino term ${\cal K}$ is invariant under
$\cG_L\times \cG_R$, but the kinetic term which includes the metric ${\cal H}_{IJ}(\mathbb{X})$ is only invariant under $\cG_L$. We recall that, on the twisted torus $\cX=\cG/\G$, only that
sub-group of $\cG_L$ which is preserved by $\G$ will have a well-defined action. In the following sections we will be particularly interested in gauging sub-groups of this rigid $\cG_L$ symmetry. Note also that the Wess-Zumino three-form ${\cal K}$ satisfies $\d{\cal K}=0$ by
virtue of the Jacobi identity $t_{[MN}{}^Qt_{P]Q}{}^T=0$. An open string version of this theory was considered in \cite{Albertsson:2008gq} and a related sigma model was investigated in \cite{Dall'Agata:2008qz}.

\subsubsection{The Constraint}

\noindent  The sigma model (\ref{fibre action}) has double the required degrees of freedom, so we seek a constraint to halve these degrees of freedom to leave the correct number. This constraint must be compatible with the equations of motion of (\ref{fibre action}) and the Maurer-Cartan equations (\ref{worldsheet MC equations}).
Under infinitesimal variations $\mathbb{X}^I\rightarrow\mathbb{X}^I+\delta \mathbb{X}^I$, the left-invariant one-forms change as ${\cal P}^M\rightarrow {\cal P}^M+\delta{\cal P}^M$, where
$$
\delta {\cal P}^M={\cal P}^M{}_I\d(\delta\mathbb{X}^I)+(\partial_J{\cal P}^M{}_I)\delta\mathbb{X}^J \d\mathbb{X}^I
$$
The equations of motion of the action (\ref{fibre action}) are then given by
\begin{eqnarray}\label{eom}
\d*{\cal M}_{MN}{\cal P}^N+\mathcal{M}_{NP}t_{MQ}{}^P{\cal P}^Q\wedge *{\cal P}^N+L_{MN}\d{\cal P}^N=0
\end{eqnarray}
These equations of motion (\ref{eom}) and the Maurer-Cartan equations
 (\ref{worldsheet MC equations}) are both consistent with
$$
\d({\cal P}^M-L^{MN}{\cal M}_{NP}*{\cal P}^P)=0
$$
and so we shall impose the   constraint \cite{New,Hull ``A geometry for non-geometric string backgrounds''}
\begin{equation}\label{constraintA}
{\cal P}^M=L^{MN}{\cal M}_{NP}*{\cal P}^P
\end{equation}
Any two of (\ref{worldsheet MC equations}), (\ref{eom}) and (\ref{constraintA}) may be used to deduce the third.

\subsubsection{The Constraint from Gauging}

\noindent The conventional space-time is recovered locally from the doubled twisted torus as a patch of the coset $\cG/\widetilde{G}_L$ where
$\widetilde{G}\subset\cG_L$ is a left acting sub-group that is also
maximally isotropic\footnote{The Lie-subalgebra is a maximally null subspace of the Lie algebra of $\widetilde{G}$
with respect to the metric $L_{MN}$ of signature $(D,D)$}.
A non-linear sigma model with target space $\cG/\widetilde{G}_L$ is obtained by gauging the left-acting $\widetilde{G}_L\subset\cG_L$ isometry sub-group of a non-linear sigma
model for the target space $\cG$ \cite{New}. The sigma model
\begin{eqnarray}\label{action}
S_{\cG}&=&\frac{1}{4}\oint_{\Sigma}{\cal H}_{IJ}\d\mathbb{X}^I\wedge*\d\mathbb{X}^J +\frac{1}{12}\int_V{\cal K}_{IJK}\d\mathbb{X}^I\wedge \d\mathbb{X}^J\wedge
\d\mathbb{X}^K
\end{eqnarray}
  has rigid $\cG_L$ symmetry, generated by the vector field
$$
\widetilde{{\cal Z}}_M=(\widetilde{{\cal P}}^{-1})_M{}^I\frac{\partial}{\partial \mathbb{X}^I}
$$
We shall be interested in gauging the null subgroup $\widetilde{G}$, which acts as $\cG\rightarrow \tilde{g}\cG$ for ${\ti g}\in\widetilde{G}$.
$\widetilde{G}_L$ is generated by the vector field $\widetilde{X}^m=\Pi^{mM}\widetilde{{\cal Z}}_M$ so that
$$
\widetilde{X}^m=\Pi^{mM}(\widetilde{{\cal P}}^{-1})_M{}^I\frac{\partial}{\partial \mathbb{X}^I}
$$
Suppose for now that $R^{mnp}=\Pi^{mM}\Pi^{nN}\Pi^{pP}t_{MNP}=0$ then the $\widetilde{X}^m$ generate a sub-group $\widetilde{G}_L$ with Lie algebra
$$
[\widetilde{X}^m,\widetilde{X}^n]=-f^{mn}{}_p\widetilde{X}^p
$$
Under the action of $\widetilde{G}_L$ the embedding fields $\XX^I$ transform infinitesimally as
$$
\delta\mathbb{X}^I=\varepsilon_m\widetilde{X}^{m}\mathbb{X}^I=\Pi^{mM}(\widetilde{{\cal P}}^{-1})_M{}^I\varepsilon_m
$$
where the parameter now depends on the world-sheet coordinates, $\varepsilon\rightarrow\varepsilon(\tau,\sigma)$. We introduce Lie algebra valued world-sheet one-forms $C_m=C_{m\alpha}\d \sigma^{\alpha}$ which transform under the gauge symmetry as
\begin{equation}\label{C trans}
\delta C_m=-\d\varepsilon_m-f^{np}{}_m\varepsilon_pC_n
\end{equation}
and define the $\widetilde{G}_L$-covariant derivatives
$$
\mathcal{D}\mathbb{X}^I=\d\mathbb{X}^I+\widetilde{X}^mC_m\mathbb{X}^I=\d\mathbb{X}^I+(\widetilde{\cal P}^{-1})_M{}^I\Pi^{Mm}C_m
$$
The kinetic term in (\ref{action}) can be made gauge invariant simply by minimal coupling giving the gauge-invariant kinetic term
\begin{equation}\label{kinetic}
S_{\text{Kin}}=\frac{1}{4}\oint_{\Sigma}{\cal H}_{IJ}\mathcal{D}\mathbb{X}^I\wedge*\mathcal{D}\mathbb{X}^J
\end{equation}
The gauging of the Wess-Zumino term is achieved following the general prescription of
 \cite{Hull:1989jk}, as described in \cite{New}.
Under an infinitesimal gauge transformation, the Wess-Zumino term changes by
$$
\delta_{\varepsilon}S_{\text{wz}}=\frac{1}{2}\int_V\delta_{\varepsilon}{\cal K}=\frac{1}{2}\oint_{\Sigma}\iota_{\varepsilon}{\cal K}
$$
where $\iota_{\varepsilon}$ is the contraction with the vector field
$\varepsilon=\varepsilon_m\widetilde{X}^m$ and can be written as $\iota_{\varepsilon}=\varepsilon_m\Pi^{mM}(\widetilde{{\cal P}}^{-1})_M{}^I\iota_I$, where $\iota_I$ is a contraction with the vector field $\partial_I$.
We have used the fact that $\d{\cal K}=0$ so that
$\delta_{\varepsilon}{\cal
K}=(\iota_{\varepsilon}\d+\d \iota_{\varepsilon}){\cal K}=\d (\iota_{\varepsilon}{\cal K})$.
One can show that
$$
\varepsilon_m\d\widetilde{P}^m=\iota_{\varepsilon}{\cal K} \qquad  \text{where}    \qquad  \widetilde{P}^m=\Pi^m{}_M\widetilde{\cal P}^M
$$
where we note that the $\widetilde{P}^m$ are globally-defined when the $\widetilde{X}^m$ are globally defined, as the two are dual to each other. The variation of the Wess-Zumino term may then be written
$$
\delta_{\varepsilon}S_{\text{wz}}=-\frac{1}{2}\oint_{\Sigma}\d\varepsilon_m\wedge \widetilde{P}^m
$$
This variation can be canceled
 by adding the term
\begin{equation}\label{SC}
S_{\text{c}}=-\frac{1}{2}\oint_{\Sigma}C_m\wedge \widetilde{P}^m
\end{equation}
where $C_m$ is the one-form transforming as (\ref{C trans}). It is not difficult to show that
$$
\delta_{\varepsilon}\widetilde{P}^m={\cal L}_{\varepsilon}\widetilde{P}^m=\varepsilon_nf^{mn}{}_p\widetilde{P}^p+L^{mn}\d\varepsilon_n
$$
where the constant $L^{mn}$ is given by
$$
L^{mn}=L_{MN}\Pi^{mM}\Pi^{nN}
$$
and we have assumed $R^{mnp}=0$, so that
$$
\delta_{\varepsilon}S_{\text{c}}=\frac{1}{2}\oint_{\Sigma}\d\varepsilon_m\wedge \widetilde{P}^m+\frac{1}{2}L^{mn}\oint_{\Sigma}C_m\wedge \d\varepsilon_n
$$
The first term in $\delta_{\varepsilon}S_{\text{c}}$ cancels the variation of the Wess-Zumino term $\delta_{\varepsilon}S_{\text{wz}}$ so that
$$
\delta_{\epsilon}(S_{\text{wz}}+S_{\text{c}})=\frac{1}{2}L^{mn}\oint_{\Sigma}C_m\wedge \d\varepsilon_n
$$
Since we require that the polarisation $\Pi^{mM}$ is null with respect to $L_{MN}$, the coefficient $L^{mn}$ vanishes and $S_{\text{wz}}+S_{\text{c}}$ is
gauge invariant. The full gauged non-linear sigma model on $\cG$ is then
$$
S_{\cG/\widetilde{G}}=\frac{1}{4}\oint_{\Sigma}{\cal H}_{IJ}\mathcal{D}\mathbb{X}^I\wedge*\mathcal{D}\mathbb{X}^J-\frac{1}{2}\oint_{\Sigma}C_m\wedge
\widetilde{P}^m+\frac{1}{12}\int_V{\cal K}_{IJK}\d\mathbb{X}^I\wedge \d\mathbb{X}^J\wedge \d\mathbb{X}^K
$$
We stress the fact that the gauging requires that $\widetilde{P}^m$ is globally defined and the gauge group $\widetilde{G}_L\subset\cG_L$ is null with respect to $L_{MN}$. We shall see explicit examples of such gaugings in the following sections.

   It is useful to define the `$\cG$-twisted' one-forms ${\cal C}=g^{-1}Cg$ so that $C_m\wedge \widetilde{P}^m=L_{MN}{\cal C}^M\wedge {\cal P}^N$. If $\widetilde{G}_L$ preserves and is preserved by $\G$, then the sigma model on $\cX$, which is also given by (\ref{action}), may be gauged and is given by
\begin{equation}\label{gauged sigma model}
S_{\cX/\widetilde{G}}=\frac{1}{4}\oint_{\Sigma}{\cal M}_{MN}{\cal P}^M\wedge *{\cal P}^N+\frac{1}{2}\oint_{\Sigma}\mathcal{C}^M\wedge *{\cal
J}_M+\frac{1}{4}\oint_{\Sigma}{\cal M}_{MN}{\cal C}^M\wedge *{\cal C}^N+\frac{1}{12}\int_Vt_{MNP}{\cal P}^M\wedge {\cal P}^N\wedge{\cal P}^P
\end{equation}
where
$$
{\cal J}_M={\cal M}_{MN}{\cal P}^N-L_{MN}*{\cal P}^N
$$
Note that the constraint (\ref{constraintA}) may be written as ${\cal J}_M=0$. More generally, $\G$ will not preserve $\widetilde{G}_L$ and we can only gauge the sigma model on a cover of $\cX$.

The conventional un-doubled
theory is recovered by integrating out the gauge fields $C_m$, which appear quadratically as auxiliary fields. Integrating out $C_m$ generates a shift in the dilaton as studied in \cite{New,Hull ``Doubled geometry and T-folds''}. Simple examples where given in \cite{New} and further examples will be studied in the following section.

\section{Polarisations and T-duality}

Once a polarisation has been chosen it is then possible to recover a conventional description of the background if one exists. The action of the discrete group $O(3,3;\Z)$ on the doubled target space $\cX$ changes the polarisation and will, in general, map one sigma model to an inequivalent
sigma model on a different background. There are circumstances in which the two models are inequivalent descriptions of the same physics but, as
argued in \cite{Giveon   ``On nonAbelian duality''}, this is generally not the case. In this section we consider the T-dualities discussed in section two from the perspective of the sigma model describing the embedding of a world-sheet into the doubled twisted torus $\cX$.

The Maurer-Cartan equations (\ref{6 dim MC}) are symmetric under the simultaneous exchange $z\leftrightarrow {\ti z}$ and
$m\leftrightarrow n$, so that a dualisation along the $z$-direction gives a dual background with the same local structure as the original,
except with the roles of $m$ and $n$ exchanged. The global structure will of course be different as the radii of the circles along the $z$
direction will be inverted by the duality.
Performing a generalised T-duality of the kind conjectured in \cite{Dabholkar ``Generalised T-duality and non-geometric backgrounds''} along the $x$-direction will also be considered. The results of this section are given in the table below:
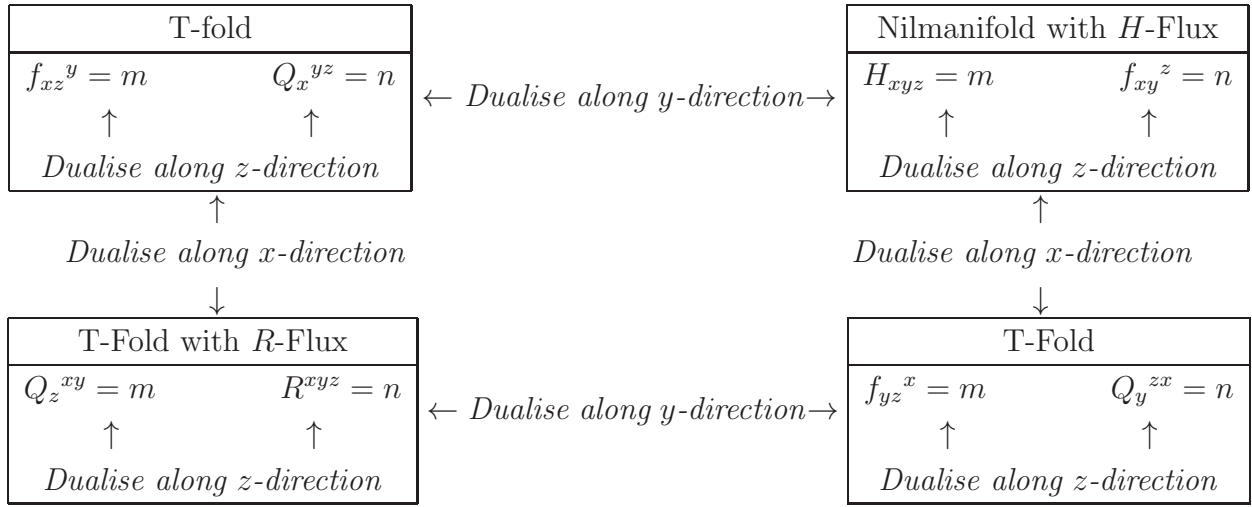
\begin{center}
\begin{tabular}{|c|}
  \hline
T-fold\\
  \hline$f_{xz}{}^y=m$\qquad\qquad$Q_x{}^{yz}=n$ \\
  $\uparrow$\qquad\qquad\qquad$\uparrow$\\
   \emph{Dualise along $z$-direction} \\
  \hline
\end{tabular}
$\leftarrow$ \emph{Dualise along $y$-direction}$\rightarrow$
\begin{tabular}{|c|}
  \hline
Nilmanifold with $H$-Flux\\
 \hline $H_{xyz}=m$\qquad\qquad$f_{xy}{}^z=n$\\
  $\uparrow$\qquad\qquad\qquad$\uparrow$\\
  \emph{Dualise along $z$-direction} \\
  \hline
\end{tabular}
\\
$\uparrow\qquad\qquad\qquad\qquad\qquad\qquad\qquad\qquad\qquad\qquad\qquad\qquad\qquad\uparrow$\\
\emph{Dualise along $x$-direction} \qquad\qquad\qquad\qquad\qquad\qquad\qquad \emph{Dualise along $x$-direction}
\\
$\downarrow\qquad\qquad\qquad\qquad\qquad\qquad\qquad\qquad\qquad\qquad\qquad\qquad\qquad\downarrow$
\\
\begin{tabular}{|c|}
  \hline
T-Fold with $R$-Flux\\
 \hline $Q_z{}^{xy}=m$\qquad\qquad$R^{xyz}=n$\\
  $\uparrow$\qquad\qquad\qquad$\uparrow$\\
  \emph{Dualise along $z$-direction} \\
  \hline
\end{tabular}
$\leftarrow$ \emph{Dualise along $y$-direction}$\rightarrow$
\begin{tabular}{|c|}
  \hline
T-Fold\\
 \hline $f_{yz}{}^x=m$\qquad\qquad$Q_y{}^{zx}=n$ \\
  $\uparrow$\qquad\qquad\qquad$\uparrow$\\
  \emph{Dualise along $z$-direction} \\
  \hline
\end{tabular}
\end{center}

The
effect of the $O(3,3;\Z)$ action is to change the subgroup $\widetilde{G}_L\subset\cG_L$ which is used to recover a local description of the internal space as a patch of $\cG/\widetilde{G}_L$. The complement of the
physical space in the double then plays the role of an auxiliary space. In all of the following examples, the coordinates on the physical internal space will be $x^i=(x,y,z)$ and the coordinates on the auxiliary space will be
$\tilde{x}_i=(\tilde{x},\tilde{y},\tilde{z})$. The labeling of coordinates will therefore depend on the polarisation chosen.

\subsection{Nilmanifold with H-Flux}

The nilmanifold background considered in section two is recovered from the doubled twisted torus by choosing the polarisation projectors
$$
\Pi=\left(
      \begin{array}{ccc}
        1 & 0 & 0 \\
        0 & 1 & 0 \\
        0 & 0 & 1 \\
        0 & 0 & 0 \\
        0 & 0 & 0 \\
        0 & 0 & 0 \\
      \end{array}
    \right)
     \qquad  \widetilde{\Pi}=\left(
                               \begin{array}{ccc}
                                 0 & 0 & 0 \\
                                 0 & 0 & 0 \\
                                 0 & 0 & 0 \\
                                 1 & 0 & 0 \\
                                 0 & 1 & 0 \\
                                 0 & 0 & 1 \\
                               \end{array}
                             \right)
$$
so that
\begin{eqnarray}
\begin{array}{lll}
 x=\Pi^x{}_I\mathbb{X}^I=\mathbb{X}^1  &\qquad  y=\Pi^y{}_I\mathbb{X}^I=\mathbb{X}^2 &\qquad z=\Pi^z{}_I\mathbb{X}^I=\mathbb{X}^3 \\
\tilde{x}=\widetilde{\Pi}_{xI}\mathbb{X}^I=\mathbb{X}^4 &\qquad  \tilde{y}=\widetilde{\Pi}_{yI}\mathbb{X}^I=\mathbb{X}^5 &\qquad
\tilde{z}=\widetilde{\Pi}_{zI}\mathbb{X}^I=\mathbb{X}^6
\end{array}
\end{eqnarray}
The structure constants which fix the local structure of
the doubled geometry in this polarisation are
\begin{equation}
\widetilde{\Pi}_x{}^M\widetilde{\Pi}_y{}^N\Pi_{zP}t_{MN}{}^P=f_{xy}{}^z=-n    \qquad  \widetilde{\Pi}_x{}^M\widetilde{\Pi}_y{}^N\widetilde{\Pi}_{zP}t_{MN}{}^P=H_{xyz}=-m
\end{equation}
The left-invariant one-forms are
\begin{eqnarray}
\begin{array}{ll}
 P^x=\d x  &\qquad  Q_x=\d\tilde{x}+my\d z+ny\d{\ti z}+mnxy\d y\\
  P^y=\d y &\qquad Q_y=\d\tilde{y}-mx\d z-nx\d{\ti z}+mnxy\d x \\
  P^z=\d z+nx\d y &\qquad Q_z=\d\tilde{z}-my\d x
\end{array}\nonumber
\end{eqnarray}
and the algebra generated by the right-invariant vector fields is
$$
 [\widetilde{Z}_y,\widetilde{Z}_z]=-m\widetilde{X}^x  \qquad  [\widetilde{Z}_x,\widetilde{Z}_z]=m\widetilde{X}^y    \qquad  [\widetilde{Z}_x,\widetilde{X}^z]=-n\widetilde{X}^y
 $$
 $$
 [\widetilde{Z}_x,\widetilde{Z}_y]=-n\widetilde{Z}_z-m\widetilde{X}^z \qquad [\widetilde{Z}_y,\widetilde{X}^z]=n\widetilde{X}^x
$$
from which one can see that the $\widetilde{Z}_m$'s do not close to form a sub-algebra. It is therefore not possible to gauge the transformations generated by the $\widetilde{Z}_m$'s in the sigma model (\ref{fibre action}), but it is possible to gauge the left-acting subgroup generated by the $\widetilde{X}^m$'s. These generators are well-defined on the group manifold $\cG$ but, under the action of $\G$, they transform as
$$
\widetilde{X}^x\rightarrow \widetilde{X}^x\qquad
\widetilde{X}^y\rightarrow \widetilde{X}^y\qquad
\widetilde{X}^z\rightarrow \widetilde{X}^z+n\beta\widetilde{X}^x-n\alpha\widetilde{X}^y
$$
for integers $\alpha,\beta$, and so are not individually well-defined on $\cX$; however, $\G$ preserves the set of generators $\widetilde{X}^m$ and so the group $\widetilde{G}_L\subset\cG_L$ is preserved by $\G$ and so we may also gauge $\widetilde{G}_L$ on the sigma model with target space $\cX$.

\subsubsection{Gauging the kinetic term}

The rigid action of the group $\widetilde{G}_L=\R^3$ leaves the space-time coordinates $x^i$ invariant but acts on the auxiliary coordinates ${\ti x}_i$ as
$$
\tilde{x}\rightarrow\tilde{x}+\tilde{\varepsilon}_x    \qquad  \tilde{y}\rightarrow\tilde{y}+\tilde{\varepsilon}_y \qquad  \tilde{z}\rightarrow\tilde{z}+\tilde{\varepsilon}_z
$$
The one-forms $P^m$ and $Q_m$ are invariant under these transformations for constant parameters $(\tilde{\varepsilon}_x,\tilde{\varepsilon}_y,\tilde{\varepsilon}_z)$ and $\widetilde{G}_L$ is a rigid symmetry of (\ref{fibre action}). If we now promote $\widetilde{G}_L$ to a local transformation and allow these parameters to depend on the world-sheet coordinates then the one-forms $P^m$ are still invariant, as they do not depend on the ${\ti x}_i$-coordinates, but the $Q_m$ transform as
$$
\delta Q_x=\d\tilde{\varepsilon}_x+ny\d\tilde{\varepsilon}_z   \qquad  \delta Q_y=\d\tilde{\varepsilon}_y-nx\d\tilde{\varepsilon}_z   \qquad  \delta Q_z=\d\tilde{\varepsilon}_z
$$
We can gauge the symmetry generated by the abelian sub-group $\widetilde{G}_L$ by introducing the world-sheet one-forms $C_m=(C_x,C_y,C_z)$ which transform as
$$
\delta C_x=-\d\tilde{\varepsilon}_x \qquad  \delta C_y=-\d\tilde{\varepsilon}_y \qquad  \delta C_z=-\d\tilde{\varepsilon}_z
$$
The gauge-invariant (minimally coupled) one forms are then
$$
{\cal P}^{\hat{M}}+{\cal C}^{\hat{M}}
$$
where ${\cal C}={\cal C}^MT_M=g^{-1}Cg$, $g\in\cG$ and
\begin{eqnarray}\label{nilfold C}
\begin{array}{ll}
 {\cal C}^x=0  &\qquad  {\cal C}_x=C_x+nyC_z\\
  {\cal C}^y=0 &\qquad {\cal C}_y=C_y-nxC_z \\
  {\cal C}^z=0 &\qquad {\cal C}_z=C_z
\end{array}
\end{eqnarray}
Since ${\cal C}^m=\Pi^m{}_M{\cal C}^M=0$ the minimally-coupled kinetic term only involves the minimal coupling $Q_m\rightarrow Q_m+{\cal C}_m$ and may be written in terms of the Lagrangian
\begin{eqnarray}\label{nilfold kinetic}
{\cal L}_{\text{Kin}}&=&\frac{1}{4}P^x\wedge *P^x+\frac{1}{4}P^y\wedge *P^y+\frac{1}{4}P^z\wedge *P^z +\frac{1}{4}(Q_x+{\cal C}_x)\wedge *(Q_x+{\cal C}_x)\nonumber\\
&&+\frac{1}{4}(Q_y+{\cal C}_y)\wedge *(Q_y+{\cal C}_y)+\frac{1}{4}(Q_z+{\cal C}_z)\wedge *(Q_z+{\cal C}_z)
\end{eqnarray}

\subsubsection{Gauging the Wess-Zumino term}

It is not hard to show that the Wess-Zumino term may be written as
$$
S_{\text{wz}}=\frac{1}{2}\oint_{\Sigma}P^m\wedge Q_m+\int_VmP^x\wedge P^y\wedge P^z
$$
and
$$
S_{\text{c}}=\frac{1}{2}\oint_{\Sigma}L_{MN}{\cal P}^M\wedge{\cal C}^N=\frac{1}{2}\oint_{\Sigma}P^m\wedge {\cal C}_m
$$
so that
\begin{equation}\label{nilfold WZW}
S_{\text{c}}+S_{\text{wz}}=\frac{1}{2}\oint_{\Sigma}P^m\wedge (Q_m+{\cal C}_m)+\int_VmP^x\wedge P^y\wedge P^z
\end{equation}

\subsubsection{Integrating out the auxiliary fields}

The action for the gauged sigma model (\ref{gauged sigma model}) may be written as
$$
S_{{\cal X}/\widetilde{G}_L}=\oint_{\Sigma}{\cal L}_{\text{Kin}}+S_{\text{c}}+S_{\text{wz}}
$$
where ${\cal L}_{\text{Kin}}$ is given by (\ref{nilfold kinetic}) and $S_{\text{c}}+S_{\text{wz}}$ is given by (\ref{nilfold WZW}). Completing the square in ${\cal C}_m$, the action splits into two parts
$$
S_{{\cal X}/\widetilde{G}_L}=S_{\cal N}[x^i]+S_{\lambda}[{\ti x}_i,C_m]
$$
where
$$
S_{\cal N}=\frac{1}{2}\oint_{\Sigma}(P^x\wedge *P^x+P^y\wedge *P^y+P^z\wedge *P^z)+\int_VmP^x\wedge P^y\wedge P^z
$$
is the action for the sigma model with target space given by the nilmanifold with constant $H$-flux and
$$
S_{\lambda}=\frac{1}{4}\oint_{\Sigma}(\lambda_x\wedge *\lambda_x+\lambda_y\wedge *\lambda_y+\lambda_z\wedge *\lambda_z)
$$
where
\begin{equation}\label{nilfold lambda}
\lambda_x={\cal C}_x+Q_x-*P^x   \qquad  \lambda_y={\cal C}_y+Q_y-*P^y   \qquad  \lambda_z={\cal C}_z+Q_z-*P^z
\end{equation}
From (\ref{nilfold C}) and (\ref{nilfold lambda}) we see that the Jacobean between the $C_m$ and $\lambda_m$ is trivial and integrating out the $C_m$ gives a nilmanifold with constant $H$-flux, $H=m\d x\wedge \d y\wedge \d z$ where the action is
$$
S_{\cal N}[x^i]=\frac{1}{2}\oint_{\Sigma}g_{ij}\d x^i\wedge *\d x^j+\int_V m\d x\wedge \d y\wedge \d z
$$
where
$$
g_{ij}=\left(%
\begin{array}{ccc}
  1 & 0 & 0 \\
  0 & 1+(nx)^2 & nx \\
  0 & nx & 1 \\
\end{array}%
\right)
$$
As shown in section two, this background has local isometries along the $y$- and $z$- directions which preserve the $H$-field. Dualising along the $z$-direction simply exchanges $m$ and $n$ and does not produce a background with a
qualitatively different local structure, although the radius of the $z$-circle will be inverted.

\subsection{T-Fold}

As demonstrated in section 2.1, it is possible to dualise the nilmanifold background along the $y$-direction by applying the Buscher rules fibre-wise, as the isometry in this direction is abelian, giving the T-fold background found in section 2.1. The effect of the
dualisation on the doubled geometry is to exchange $P^y$ with $Q_y$, $y$ with $\tilde{y}$ and $\widetilde{Z}_y$ with $\widetilde{X}^y$. The T-fold polarisation is recovered from the doubled twisted torus by choosing the polarisation projectors
\begin{equation}\label{T-fold pol}
\Pi=\left(
      \begin{array}{ccc}
        1 & 0 & 0 \\
        0 & 0 & 0 \\
        0 & 0 & 1 \\
        0 & 0 & 0 \\
        0 & 1 & 0 \\
        0 & 0 & 0 \\
      \end{array}
    \right)
     \qquad  \widetilde{\Pi}=\left(
                               \begin{array}{ccc}
                                 0 & 0 & 0 \\
                                 0 & 1 & 0 \\
                                 0 & 0 & 0 \\
                                 1 & 0 & 0 \\
                                 0 & 0 & 0 \\
                                 0 & 0 & 1 \\
                               \end{array}
                             \right)
\end{equation}
so that locally we can identify the coordinates
\begin{eqnarray}
\begin{array}{lll}
 x=\Pi^x{}_I\mathbb{X}^I=\mathbb{X}^1  &\qquad  y=\Pi^y{}_I\mathbb{X}^I=\mathbb{X}^5 &\qquad z=\Pi^z{}_I\mathbb{X}^I=\mathbb{X}^3 \\
\tilde{x}=\widetilde{\Pi}_{xI}\mathbb{X}^I=\mathbb{X}^4 &\qquad  \tilde{y}=\widetilde{\Pi}_{yI}\mathbb{X}^I=\mathbb{X}^2 &\qquad
\tilde{z}=\widetilde{\Pi}_{zI}\mathbb{X}^I=\mathbb{X}^6
\end{array}
\end{eqnarray}
The structure constants which determine
the doubled geometry in this polarisation are
\begin{equation}
\widetilde{\Pi}_x{}^M\Pi^{yN}\Pi^z{}_Pt_{MN}{}^P=Q_x{}^{yz}=-n    \qquad  \widetilde{\Pi}_x{}^M\widetilde{\Pi}_z{}^N\Pi^y{}_Pt_{MN}{}^P=f_{xz}{}^y=m
\end{equation}
The left-invariant one-forms are
\begin{eqnarray}
\begin{array}{ll}
 P^x=\d x  &\qquad  Q_x=\d\tilde{x}+m\tilde{y}\d z+n\tilde{y}\d{\ti z}+mn\tilde{y}x\d{\ti y}\\
  P^y=\d y-mx\d z-nx\d{\ti z}+mnx{\ti y}\d x &\qquad Q_y=\d\tilde{y} \\
  P^z=\d z+nx\d{\ti y} &\qquad Q_z=\d\tilde{z}-m\tilde{y}\d x
\end{array}\nonumber
\end{eqnarray}
and the Lie algebra generated by the right-invariant vector fields is
$$
 [\widetilde{X}^y,\widetilde{Z}_z]=-m\widetilde{X}^x  \qquad  [\widetilde{Z}_x,\widetilde{Z}_z]=m\widetilde{Z}_y    \qquad  [\widetilde{Z}_x,\widetilde{X}^z]=-n\widetilde{Z}_y
 $$
 $$
 [\widetilde{Z}_x,\widetilde{X}^y]=-n\widetilde{Z}_z-m\widetilde{X}^z \qquad [\widetilde{X}^y,\widetilde{X}^z]=n\widetilde{X}^x
$$
It is interesting to note that the above algebra is a Drinfel'd double \cite{New}, where both sub-groups, $G_L$ (generated by $\widetilde{Z}_m$) and $\widetilde{G}_L$ (generated by $\widetilde{X}^m$), are Heisenberg groups. We now consider the gauging of the sigma model in this polarisation. The polarisation selects the generators $(\widetilde{X}^x,\widetilde{X}^y,\widetilde{X}^z)=(\widetilde{\cal Z}_4,\widetilde{\cal Z}_2,\widetilde{\cal Z}_6)$ which close to generate the three-dimensional non-abelian group $\widetilde{G}_L$. These generators are well-defined on the doubled group $\cG$ but under the action of $\G$, the they transform as
\begin{eqnarray}
\widetilde{X}^x&\rightarrow &\widetilde{X}^x\nonumber\\
\widetilde{X}^y&\rightarrow& \widetilde{X}^y-n\alpha\widetilde{Z}_z-m\alpha\widetilde{X}^z -mn\alpha^2\widetilde{Z}_y +(m\gamma+n\tilde{\gamma}-mn\alpha\beta)\widetilde{X}^x\nonumber\\
\widetilde{X}^z&\rightarrow& \widetilde{X}^z+n\beta\widetilde{X}^x-n\alpha\widetilde{Z}_y
\end{eqnarray}
for constant integers $(\alpha,\gamma,\tilde{\beta},\tilde{\gamma})$, and we see that the $\widetilde{X}^m$'s mix with the $\widetilde{Z}_m$'s mix together and $\widetilde{G}_L$ is not preserved by $\G$ and so the generators $\widetilde{X}^m$ are not well-defined on the doubled twisted torus $\cX$. However, if we take the cover, ${\cal C}_{\cal X}$, given by replacing $S^1_x$ with $\R_x$ (and simply continuing the fields on $\cX$ in $x$) we see that the set of generators of $\widetilde{G}_L$ are preserved by the action of $\G'$
\begin{eqnarray}
\widetilde{X}^x&\rightarrow &\widetilde{X}^x\nonumber\\
\widetilde{X}^y&\rightarrow& \widetilde{X}^y+(m\gamma+n\tilde{\gamma})\widetilde{X}^x\nonumber\\
\widetilde{X}^z&\rightarrow& \widetilde{X}^z+n\beta\widetilde{X}^x
\end{eqnarray}
where $\G'$ is the sub-group of $\G$ that does not impose the identification $x\sim x+1$. We may then gauge the subgroup $\widetilde{G}_L$ of the sigma model with target space ${\cal C}_{\cal X}$ to give a sigma model with target space ${\cal C}_{\cal X}/\widetilde{G}_L$

\subsubsection{Gauging the kinetic term}

A more convenient parametrisation of the doubled group is given by the coordinate redefinitions
\begin{equation}\label{coordinate transfromations}
{\ti x}\rightarrow {\ti x}'={\ti x}-mz\tilde{y}-mnx\tilde{y}^2  \qquad  \tilde{z}\rightarrow\tilde{z}'=\tilde{z}+mx\tilde{y}
\end{equation}
so that the left-invariant forms on $\cX$ may be written as
\begin{eqnarray}\label{forms}
\begin{array}{ll}
 P^x=\d x  &\qquad  Q_x=\d\tilde{x}+n\tilde{y}\d{\ti z}-mz\d {\ti y}\\
  P^y=\d y-mx\d z-nx\d{\ti z}-mnx^2\d{\ti y} &\qquad Q_y=\d\tilde{y} \\
  P^z=\d z+nx\d{\ti y} &\qquad Q_z=\d\tilde{z}+mx\d\tilde{y}
\end{array}
\end{eqnarray}
The subgroup $\widetilde{G}_L$ generated by $\widetilde{X}^m$ does not act on the coordinates $(x,y,x)$ but acts on the coordinates $(\tilde{x},\tilde{y},\tilde{z})$ as
$$
\tilde{x}\rightarrow \tilde{x}-n\tilde{\varepsilon}_y\tilde{z}+\tilde{\varepsilon}_x   \qquad  \tilde{y}\rightarrow\tilde{y}+\tilde{\varepsilon}_y \qquad  \tilde{z}\rightarrow\tilde{z}+\tilde{\varepsilon}_z
$$
where $(\tilde{\varepsilon}_x,\tilde{\varepsilon}_y,\tilde{\varepsilon}_z)$ are parameters of the sub-group $\widetilde{G}_L$. Unlike the previous nilmanifold example, the sub-group $\widetilde{G}_L$ is non-abelian with structure constant $f_x{}^{yz}=n\in \Z$. For constant $(\tilde{\varepsilon}_x,\tilde{\varepsilon}_y,\tilde{\varepsilon}_z)$ the left-invariant one-forms (\ref{forms}) are preserved by the action of $\widetilde{G}_L$ but we are interested in gauging this symmetry and if we let these parameters depend on the world-sheet coordinates then the one-forms $Q_m$ transform under the infinitesimal action of $\widetilde{G}_L$ as
$$
\delta Q_x=\d\tilde{\varepsilon}_x-(n\tilde{z}+mz)\d\tilde{\varepsilon}_y+n\tilde{y}\d\tilde{\varepsilon}_z  \qquad  \delta Q_y=\d\tilde{\varepsilon}_y   \qquad  \delta Q_z=\d\tilde{\varepsilon}_z+mx\d\tilde{\varepsilon}_y
$$
We can now appreciate the utility of the coordinate redefinitions (\ref{coordinate transfromations}) which are such that the gauge parameters only appear in the infinitesimal variations of $Q_m$ under a derivative as $\d\tilde{\varepsilon}_m$. In order to minimally couple the kinetic term in (\ref{action}) we introduce the world-sheet one-forms $C_m=(C_x,C_y,C_z)$ which transform under the local non-abelian $\widetilde{G}_L$ as
$$
\delta C_x=-\d\tilde{\varepsilon}_x-n\tilde{\varepsilon}_yC_z+n\tilde{\varepsilon}_zC_y \qquad  \delta C_y=-\d\tilde{\varepsilon}_y \qquad  \delta C_z=-\d\tilde{\varepsilon}_z
$$
As for the nilmanifold case, it is useful to define the $\cG$-twisted one-forms ${\cal C}=g^{-1}Cg$, where
\begin{eqnarray}\label{minimally coupled forms}
\begin{array}{ll}
 {\cal C}^x=0  &\qquad  {\cal C}_x=C_x-(n\tilde{z}+mz)C_y+n\tilde{y}C_z\\
  {\cal C}^y=-nx(C_z+mxC_y) &\qquad {\cal C}_y=C_y \\
  {\cal C}^z=nxC_y &\qquad {\cal C}_z=C_z+mxC_y
\end{array}\nonumber
\end{eqnarray}
The gauge-invariant (minimally coupled) one-forms are then ${\cal P}^{\hat{M}}+{\cal C}^{\hat{M}}$, the components of which are
\begin{eqnarray}
\begin{array}{ll}
 P^x+{\cal C}^x=\ell^x  &\qquad  Q_x+{\cal C}_x=Q_x+C_x-(n\tilde{z}+mz)C_y+n\tilde{y}C_z\\
 P^y+ {\cal C}^y=\ell^y-nx(Q_z+{\cal C}_z) &\qquad Q_y+{\cal C}_y=Q_y+C_y \\
 P^z+ {\cal C}^z=\ell^z+nx(Q_y+{\cal C}_y) &\qquad Q_z+{\cal C}_z=Q_z+C_z+mxC_y
\end{array}\nonumber
\end{eqnarray}
where it is useful to define
\begin{equation}\label{L's}
\ell^x=\d x   \qquad  \ell^y=\d y-mx\d z  \qquad  \ell^z=\d z
\end{equation}
which are the left-invariant one-forms of a Heisenberg group manifold. It is helpful to write the minimally coupled left-invariant one forms as
$$
{\cal P}^{\hat{M}}+{\cal C}^{\hat{M}}=\Phi^{\hat{N}}{\cal V}_{\hat{N}}{}^{\hat{M}}
$$
where $\Phi^{\hat{M}}=(\ell^m,Q_m+\mathcal{C}_m)$ and
$$
{\cal V}=\left(
           \begin{array}{cc}
             1_3 & 0  \\
             \beta & 1_3 \\
           \end{array}
         \right)    \qquad  \text{where}    \qquad  \beta=\left(
           \begin{array}{ccc}
             0 & 0 & 0 \\
             0 & 0 & nx \\
             0 & -nx & 0 \\
           \end{array}
         \right)
$$
The minimally-coupled kinetic term (\ref{kinetic}) is then written as
$$
S_{\text{Kin}}=\frac{1}{4}\oint_{\Sigma} \delta_{MN}({\cal P}^M+{\cal C}^M)\wedge*({\cal P}^N+{\cal C}^N) =\frac{1}{4}\oint_{\Sigma} ({\cal V}{\cal V}^t)_{\hat{M}\hat{N}}\Phi^{\hat{M}}\wedge*\Phi^{\hat{N}}
$$
where we have taken ${\cal M}_{MN}=\delta_{MN}$ and $({\cal V}{\cal V}^t)_{\hat{M}\hat{N}}={\cal V}_{\hat{M}}{}^{\hat{P}}\delta_{\hat{P}\hat{Q}}{\cal V}^{\hat{Q}}{}_{\hat{N}}$. The kinetic term may then be given in terms of the Lagrangian ${\cal L}_{\text{Kin}}$ which is written in terms of the physical space-time coordinates $x^i$ and the gauge-invariant one-forms $Q_m+\mathcal{C}_m$
\begin{eqnarray}\label{T-fold kinetic}
{\cal L}_{\text{Kin}}&=&\frac{1}{4}\ell^x\wedge *\ell^x+\frac{1}{4}\ell^y\wedge *\ell^y+\frac{1}{4}\ell^z\wedge *\ell^z +\frac{1}{4}(Q_x+\mathcal{C}_x)\wedge *(Q_x+\mathcal{C}_x)\nonumber\\
&&+ \frac{1+(nx)^2}{4}(Q_y+\mathcal{C}_y)\wedge *(Q_y+\mathcal{C}_y)+ \frac{1+(nx)^2}{4}(Q_z+\mathcal{C}_z)\wedge *(Q_z+\mathcal{C}_z)\nonumber\\
&&-\frac{nx}{2}\ell^y\wedge *(Q_z+\mathcal{C}_z)+\frac{nx}{2}\ell^z\wedge *(Q_y+\mathcal{C}_y)
\end{eqnarray}

\subsubsection{Gauging the Wess-Zumino term}

The Wess-Zumino term for the doubled sigma model (\ref{action}) can be written
\begin{eqnarray}
S_{\text{wz}}&=&\frac{1}{2}\int_V\left(mP^x\wedge P^z\wedge Q_y-nP^x\wedge Q_y\wedge Q_z\right)\nonumber\\
&=&\frac{1}{2}\oint_{\Sigma}\ell^m\wedge Q_m\nonumber
\end{eqnarray}
where the $\ell^m$ are given by (\ref{L's}). In this polarisation, the Lagrangian for $S_c$ (\ref{SC}) can be written
\begin{equation}\label{T-fold SC}
{\cal L}_{\text{c}}=\frac{1}{2}L_{MN}{\cal P}^{M}\wedge {\cal C}^N=\frac{1}{2}\ell^m\wedge {\cal C}_m
\end{equation}
so that
\begin{equation}\label{T-fold WZW}
{\cal L}_{\text{c}}+{\cal L}_{\text{wz}}=\frac{1}{2}\ell^m\wedge(Q_m+{\cal C}_m)
\end{equation}
is manifestly gauge-invariant.

\subsubsection{Integrating out the auxiliary fields}

In the T-fold polarisation (\ref{T-fold pol}), the gauged sigma model (\ref{gauged sigma model}) may be written in terms of the Lagrangian on the two-dimensional world-sheet
$$
{\cal L}_{{\cal C}_{\cal X}/\widetilde{G}_L}={\cal L}_{\text{Kin}}+{\cal L}_{\text{c}}+{\cal L}_{\text{wz}}
$$
where ${\cal L}_{\text{Kin}}$ is given by (\ref{T-fold kinetic}) and ${\cal L}_{\text{c}}+{\cal L}_{\text{wz}}$ is given by (\ref{T-fold WZW}). Completing the square in $Q_m+{\cal C}_m$, the action then splits into two parts
$$
S_{{\cal C}_{\cal X}/\widetilde{G}_L}[\XX^I,C_m]=S_T[x^i]+S_{\lambda}[x,{\ti x}_i,C_m]
$$
where
$$
S_T=\frac{1}{2}\oint_{\Sigma}\ell^x\wedge *\ell^x+\frac{1}{2}\oint_{\Sigma}\frac{1}{1+(nx)^2}(\ell^y\wedge *\ell^y+\ell^z\wedge *\ell^z) -\oint_{\Sigma}\frac{nx}{1+(nx)^2}\ell^y\wedge \ell^z
$$
and
$$
S_{\lambda}=\frac{1}{4}\oint_{\Sigma}\lambda_x\wedge *\lambda_x+\frac{1}{4}\oint_{\Sigma}(1+(nx)^2)\left(\lambda_y\wedge *\lambda_y+\lambda_z\wedge *\lambda_z\right)
$$
where
\begin{eqnarray}\label{T-fold lambda}
\lambda_x&=&Q_x+C_x-(n\tilde{z}+mz)C_y+n\tilde{y}C_z-*\ell^x\nonumber\\
\lambda_y&=&Q_y+C_y-\frac{1}{1+(nx)^2}(*\ell^y-nx\ell^z)\nonumber\\
\lambda_z&=&Q_z+C_z+mxC_y-\frac{1}{1+(nx)^2}(*\ell^z+nx\ell^y)
\end{eqnarray}
From the expressions (\ref{T-fold lambda}), the Jacobean between the $C_m$ and $\lambda_m$ is found to be trivial so that integrating out the $C_m$ in the path integral just leads to a determinant factor which can be absorbed into a shift of the dilaton
$$
\phi\rightarrow\phi-\ln(1+(nx)^2)
$$
We are then left with the action for a sigma model on ${\cal C}_{\cX}/\widetilde{G}_L$
$$
S_T[x^i]=\frac{1}{2}\oint_{\Sigma}g_{ij}\d x^i\wedge*\d x^j+\frac{1}{2}\oint_{\Sigma}B_{ij}\d x^i\wedge \d x^j
$$
where
\begin{eqnarray}
g_{ij}=\frac{1}{1+(nx)^2}\left(%
\begin{array}{ccc}
  1+(nx)^2 & 0 & 0 \\
  0 & 1 & -mx \\
  0 & -mx & 1+(mx)^2 \\
\end{array}%
\right) \qquad  B_{ij}=\frac{1}{1+(nx)^2}\left(%
\begin{array}{ccc}
  0 & 0 & 0 \\
  0 & 0 & nx \\
  0 & -nx & 0 \\
\end{array}%
\right)\nonumber
\end{eqnarray}
which has isometries in the $y$ and $z$ directions. This background is a conventional Riemannian geometry which is compact in the $y$- and $z$-directions and non-compact in the $x$-direction and is topologically $T^2\times S^1$. This is only a cover of the background we are really interested in. To recover the T-fold background of section 2.1 we must impose the identification $x\sim x+1$. The background is now only a conventional Riemannian geometry in patches and globally is the T-fold of section 2.1 as expected. Dualising along the $z$-direction simply gives a background in which the roles of $m$
and $n$ are exchanged and the radius of the compact $z$-direction is inverted.

\subsection{T-fold with R-Flux}

We now consider an application of the proposed generalised (non-isometric) duality of \cite{Dabholkar ``Generalised T-duality and non-geometric backgrounds''}. This is achieved by acting with an element of $O(3,3;\Z)$ on the doubled geometry, exchanging the $x$- and $\ti x$-directions with respect to the T-fold polarisation and is equivalent to choosing the polarisation
$$
\Pi=\left(
      \begin{array}{ccc}
        0 & 0 & 0 \\
        0 & 0 & 0 \\
        0 & 0 & 1 \\
        1 & 0 & 0 \\
        0 & 1 & 0 \\
        0 & 0 & 0 \\
      \end{array}
    \right)
     \qquad  \widetilde{\Pi}=\left(
                               \begin{array}{ccc}
                                 1 & 0 & 0 \\
                                 0 & 1 & 0 \\
                                 0 & 0 & 0 \\
                                 0 & 0 & 0 \\
                                 0 & 0 & 0 \\
                                 0 & 0 & 1 \\
                               \end{array}
                             \right)
$$
so that
\begin{eqnarray}
\begin{array}{lll}
 x=\Pi^x{}_I\mathbb{X}^I=\mathbb{X}^4  &\qquad  y=\Pi^y{}_I\mathbb{X}^I=\mathbb{X}^5 &\qquad z=\Pi^z{}_I\mathbb{X}^I=\mathbb{X}^3 \\
\tilde{x}=\widetilde{\Pi}_{xI}\mathbb{X}^I=\mathbb{X}^1 &\qquad  \tilde{y}=\widetilde{\Pi}_{yI}\mathbb{X}^I=\mathbb{X}^2 &\qquad
\tilde{z}=\widetilde{\Pi}_{zI}\mathbb{X}^I=\mathbb{X}^6
\end{array}
\end{eqnarray}
The initial stages of the analysis of the background given by this choice of polarisation proceed in much the same way as in the nilmanifold and T-fold examples studied above. For example, the structure constants which determine
the description of the doubled geometry in this polarisation are
$$
\Pi^{xM}\Pi^{yN}\Pi^z{}_Pt_{MN}{}^P=R^{xyz}=-n    \qquad  \Pi^{xM}\widetilde{\Pi}_z{}^N\Pi^y{}_Pt_{MN}{}^P=Q_z{}^{xy}=m
$$
and the left-invariant one-forms on $\cX$ are;
\begin{eqnarray}\label{R-flux forms}
\begin{array}{ll}
 P^x=\d x+m\tilde{y}\d z+n\tilde{y}\d{\ti z}+mn\tilde{x}\tilde{y}\d\tilde{y}  &\qquad  Q_x=\d\tilde{x}\\
  P^y=\d y-m\tilde{x}\d z-n\tilde{x}\d{\ti z}+mn\tilde{x}\tilde{y}\d\tilde{x} &\qquad Q_y=\d\tilde{y} \\
  P^z=\d z+n\tilde{x}\d\tilde{y} &\qquad Q_z=\d\tilde{z}-my\d x
\end{array}
\end{eqnarray}
however, if we now consider the Lie algebra generated by the right-invariant vector fields
$$
 [\widetilde{X}^y,\widetilde{Z}_z]=-m\widetilde{Z}_x  \qquad  [\widetilde{X}^x,\widetilde{Z}_z]=m\widetilde{Z}_y    \qquad  [\widetilde{X}^x,\widetilde{X}^z]=-n\widetilde{Z}_y
 $$
 $$
 [\widetilde{X}^x,\widetilde{X}^y]=-n\widetilde{Z}_z-m\widetilde{X}^z \qquad [\widetilde{X}^y,\widetilde{X}^z]=n\widetilde{Z}_x
$$
we see that this polarisation is of a qualitatively different type of background to the two previous examples. Specifically, we see that the generators $\widetilde{X}^m$ do not close to form a sub-algebra and so this background corresponds to the Type III scenario discussed in section two, where the obstruction to the closure of the $\widetilde{X}^m$ generators under the Lie bracket is the $R$-flux $R^{xyz}=n$. In such cases a conventional description of the background, in terms of a three-dimensional Riemannian manifold, cannot be found even locally.

Without a subgroup $\widetilde{G}_L$, generated by $\widetilde{X}^m$, it does not make sense to talk of gauging the doubled sigma model (\ref{action}) to recover a conventional sigma model on this background; however, it is possible to recover a classical description of this background by eliminating the $\d\tilde{x}_i$ dependence in the equations of motion as demonstrated in \cite{New}. Taking ${\cal M}_{MN}=\delta_{MN}$, the self-duality constraint (\ref{constraintA}) is
$$
Q_m=\delta_{mn}*P^n
$$
which relates the equations of motion (\ref{eom}) to the Maurer-Cartan equations (\ref{worldsheet MC equations}). For example, the condition $\d Q_x=0$ gives the $x$ equation of motion $\d*P^x=0$. We would like to rewrite this equation of motion, and the equations of motion for $y$ and $z$, in terms of $\tilde{x}_i$ and $\d x^i$ only by using the self-duality constraint to eliminate all $\d{\ti x}_i$-dependence in the equations of motion. It is useful to write the expressions (\ref{R-flux forms}), in matrix notation, as
$$
\textsf{A} *\textsf{Q}=\d\textsf{x}+\textsf{B} \textsf{Q}
$$
where $\textsf{A}$ and $\textsf{B}$ are $3\times 3$ matrices and $\textsf{Q}$ is a 3-vector with components $Q_m$. \textsf{A} is always invertible, but \textsf{B} may not be (for example if $m$ or $n$ are zero). The $Q_m$ may be given by
\begin{equation}\label{R-flux Q}
\textsf{Q}=\frac{1}{\textsf{1}-\textsf{A}^{-1}\textsf{B}\textsf{A}^{-1}\textsf{B}}(\textsf{A}^{-1}*\d\textsf{x} +\textsf{A}^{-1}\textsf{B}\textsf{A}^{-1}\d\textsf{x})
\end{equation}
Wherever possible, we shall try to find local coordinates for the doubled group such that $\textsf{A}=1$ and $\textsf{B}$ is antisymmetric as the calculations then simplify considerably. In particular, in such cases the expression for $\textsf{Q}$ becomes
\begin{equation}\label{Q=...}
\textsf{Q}=\frac{1}{\textsf{1}-\textsf{B}^2}(*\d\textsf{x} +\textsf{B}\d\textsf{x})
\end{equation}
We do not consider the full problem here but instead consider two illustrative examples:

\noindent\underline{$n=0$}

\noindent If we set $n=0$, the left-invariant one-forms (\ref{R-flux forms}) become
\begin{eqnarray}
\begin{array}{lll}
 P^x=\d x+m\tilde{y}\d z  &\qquad   P^y=\d y-m\tilde{x}\d z  &\qquad P^z=\d z\\
Q_x=\d\tilde{x}   &\qquad Q_y=\d\tilde{y} &\qquad Q_z=\d\tilde{z}-my\d x
\end{array}\nonumber
\end{eqnarray}
Using the reparametrisation
\begin{equation}\label{reperam}
x\rightarrow x'=x-mz{\ti y} \qquad  y\rightarrow y'=y+m{\ti x}z
\end{equation}
these left-invariant one-forms become
\begin{eqnarray}
\begin{array}{lll}
 P^x=\d x-mz\d\tilde{y}  &\qquad  P^y=\d y+mz\d\tilde{x}   &\qquad P^z=\d z\\
Q_x=\d\tilde{x}   &\qquad Q_y=\d\tilde{y} &\qquad Q_z=\d\tilde{z}-my\d x
\end{array}\nonumber
\end{eqnarray}
so that the matrices $\textsf{A}$ and $\textsf{B}$ in (\ref{R-flux Q}) and (\ref{Q=...}) are given by
$$
\textsf{A}=\left(
             \begin{array}{ccc}
               1 & 0 & 0 \\
               0 & 1 & 0 \\
               0 & 0 & 1 \\
             \end{array}
           \right)  \qquad  \textsf{B}=\left(
                                         \begin{array}{ccc}
                                           0 & -mz & 0 \\
                                           mz & 0 & 0 \\
                                           0 & 0 & 0 \\
                                         \end{array}
                                       \right)
$$
The coordinate reparametrisation (\ref{reperam}) was chosen so that $\textsf{A}=1$ and $\textsf{B}$ is anti-symmetric. Using these matrices in (\ref{Q=...}) gives the following expressions for $\textsf{Q}$
$$
Q_x=\frac{1}{1+(mz)^2}(*\d x-mz\d y)\qquad
Q_y=\frac{1}{1+(mz)^2}(*\d y+mz\d x)\qquad Q_z=*\d z
$$
The Maurer-Cartan equations
$$
\d Q_x=0  \qquad \d Q_y=0   \qquad  \d Q_z+mQ_y\wedge Q_x=0
$$
then give the equations of motion
\begin{eqnarray}
\d\left(\frac{1}{1+(mz)^2}(*\d x-mz\d y)\right)=0\qquad
\d\left(\frac{1}{1+(mz)^2}(*\d y+mz\d x)\right)=0\nonumber\\
\d*\d z+\frac{m(1-mz)^2}{(1+(mz)^2)^2}\d x\wedge \d y+\frac{m^2z}{(1+(mz)^2)^2}(\d x\wedge *\d x+\d y\wedge*\d y)=0\nonumber
\end{eqnarray}
which, in turn, may be derived from the world-sheet Lagrangian
$$
{\cal L}=\frac{1}{2}\d z\wedge*\d z+\frac{1}{2(1+(mz)^2)}(\d x\wedge*\d x+\d y\wedge*\d y)+\frac{mz}{1+(mz)^2}\d x\wedge \d y
$$
This is simply the Lagrangian for a world-sheet embedding into a T-fold given by a $T^2$, with coordinates $x$ and $y$, fibred over a base circle $S^1_z$ with a monodromy in $O(2,2;\Z)$ that includes strict T-dualities along the $x$- and $y$-directions.

\noindent\underline{$m=0$}

\noindent Next we consider the complimentary example where $m=0$. In this case the left-invariant one-forms (\ref{R-flux forms}) become
\begin{eqnarray}
\begin{array}{lll}
 P^x=\d x+n\tilde{y}\d{\ti z}  &\qquad P^y=\d y-n\tilde{x}\d{\ti z}  &\qquad P^z=\d z+n\tilde{x}\d\tilde{y}\\
Q_x=\d\tilde{x}   &\qquad Q_y=\d\tilde{y}  &\qquad Q_z=\d\tilde{z}
\end{array}\nonumber
\end{eqnarray}
The reperametrisation
$$
x\rightarrow x'=x-\frac{1}{2}n\tilde{y}\tilde{z} \qquad  y\rightarrow y'=y+\frac{1}{2}n\tilde{z}\tilde{x} \qquad  z\rightarrow z'=z-\frac{1}{2}n\tilde{x}\tilde{y}
$$
can be used to ensure that $\textsf{A}=1$ and $\textsf{B}$ is antisymmetric. In terms of these new coordinates, the left-invariant one-forms on $\cX$ in this polarisation are
\begin{eqnarray}
\begin{array}{lll}
 P^x=\d x+\frac{1}{2}n\tilde{y}\d{\ti z}-\frac{1}{2}n{\ti z}\d\tilde{y}  &\qquad P^y=\d y+\frac{1}{2}n\tilde{z}\d{\ti x}-\frac{1}{2}n{\ti x}\d\tilde{z}  &\qquad P^z=\d z+\frac{1}{2}n\tilde{x}\d{\ti y}-\frac{1}{2}y{\ti z}\d\tilde{x}\\
Q_x=\d\tilde{x}   &\qquad Q_y=\d\tilde{y}  &\qquad Q_z=\d\tilde{z}
\end{array}\nonumber
\end{eqnarray}
and the matrices $\textsf{A}$ and $\textsf{B}$ in (\ref{R-flux Q}) and (\ref{Q=...}) are given by
$$
\textsf{A}=\left(
             \begin{array}{ccc}
               1 & 0 & 0 \\
               0 & 1 & 0 \\
               0 & 0 & 1 \\
             \end{array}
           \right)  \qquad  \textsf{B}=\frac{1}{2}\left(
                                         \begin{array}{ccc}
                                           0 & -n{\ti z} & n{\ti y} \\
                                           n{\ti z} & 0 & -n{\ti x} \\
                                           -n{\ti y} & n{\ti x} & 0 \\
                                         \end{array}
                                       \right)
$$
with a little work, one can then find expressions for $\textsf{Q}$ given in (\ref{R-flux Q}) and (\ref{Q=...})
\begin{eqnarray}
Q_x&=&\frac{1}{T}\left(\zeta_x*\d x+\frac{1}{2}n({\ti z}\d y+{\ti y}\d z)+\frac{1}{4}n^2({\ti x}{\ti y}*\d y+{\ti x}{\ti z}*\d z)\right)\nonumber\\
Q_y&=&\frac{1}{T}\left(\zeta_y*\d y+\frac{1}{2}n({\ti x}\d z+{\ti z}\d x)+\frac{1}{4}n^2({\ti y}{\ti z}*\d z+{\ti y}{\ti x}*\d x)\right)\nonumber\\
Q_x&=&\frac{1}{T}\left(\zeta_z*\d z+\frac{1}{2}n({\ti y}\d x+{\ti x}\d y)+\frac{1}{4}n^2({\ti z}{\ti x}*\d x+{\ti z}{\ti y}*\d y)\right)\nonumber
\end{eqnarray}
where it is useful to define
$$
\zeta_x=1+\frac{1}{4}(n{\ti x})^2   \qquad  \zeta_y=1+\frac{1}{4}(n{\ti y})^2   \qquad  \zeta_z=1+\frac{1}{4}(n{\ti z})^2
$$
$$
T=1+\frac{1}{4}n^2({\ti x}^2+{\ti y}^2+{\ti z}^2)
$$
The equations of motion, which come from $\d Q_x=0$, $\d Q_y=0$ and $\d Q_z=0$, are given by the world-sheet action
$$
S=\frac{1}{2}\oint_{\Sigma}g_{ij}\d x^i\wedge*\d x^j+\frac{1}{2}\oint_{\Sigma}B_{ij}\d x^i\wedge \d x^j
$$
where one can define an effective metric and $B$-field for the classical background in this polarisation
$$
g_{ij}=\frac{1}{4T}\left(
         \begin{array}{ccc}
           4\zeta_x & n^2{\ti x}{\ti y} & n^2{\ti x}{\ti z} \\
           n^2{\ti x}{\ti y} & 4\zeta_y & n^2{\ti y}{\ti z} \\
           n^2{\ti x}{\ti z} & n^2{\ti y}{\ti z} & 4\zeta_z \\
         \end{array}
       \right)  \qquad  B_{ij}=\frac{1}{2T}\left(
                                 \begin{array}{ccc}
                                   0 & -n{\ti z} & n{\ti y} \\
                                   n{\ti z} & 0 & -n{\ti x} \\
                                   -n{\ti y} & n{\ti x} & 0 \\
                                 \end{array}
                               \right)
$$
The strange appearance of the coordinates conjugate to winding modes in these effective background fields may seem startling but, as commented on in \cite{New,Dabholkar ``Generalised T-duality and non-geometric backgrounds''}, it is not without precedent. A similar phenomenon was observed in \cite{Gregory:1997te,Tong:2002rq,Harvey:2005ab} where a non-trivial dependence on the dual coordinates emerged from an analysis of world-sheet instanton effects. This apparent connection between $R$-flux backgrounds, non-isometric T-duality and world-sheet instanton effects is suggestive of a more general class of phenomena and is currently under investigation.

The solution to the more general problem where $m,n\neq 0$ will be locally non-geometric and is probably best described as a T-fold with $R$-flux.

\section{Recovering the doubled torus bundle}

In this section we make contact with the doubled torus formalism of \cite{Hull ``A geometry for non-geometric string backgrounds''} by showing how a partial polarisation - a null polarisation which selects a sub-group $\widetilde{G}_L$ of dimension less than three - can be used to recover a five-dimensional doubled torus bundle ${\cal T}$ of the kind reviewed in the Introduction. We introduce the polarisation $\Pi^x{}_M$ and its complement $\widetilde{\Pi}_{xM}$ such that
$$
\Pi^x{}_M{\cal P}^M=P^x \qquad  \widetilde{\Pi}_{xM}{\cal P}^M=Q_x
$$
where
$$
\Pi^x{}_M=\left(
            \begin{array}{cccccc}
              1 & 0 & 0 & 0 & 0 & 0 \\
            \end{array}
          \right)   \qquad  \widetilde{\Pi}_{xM}=\left(
            \begin{array}{cccccc}
              0 & 0 & 0 & 1 & 0 & 0 \\
            \end{array}
          \right)
$$
and similarly for the coordinates $\Pi^x{}_I\mathbb{X}^I=x$, $\Pi_{xI}\mathbb{X}^I=\tilde{x}$. The left-invariant one-forms may then be written as
\begin{eqnarray}
\begin{array}{ll}
P^x=\d x  &\qquad  Q_x=\d{\ti x}+m\mathbb{X}^2\d\mathbb{X}^3+n\mathbb{X}^2\d\mathbb{X}^6+mnx\mathbb{X}^2\d\mathbb{X}^2  \\
{\cal P}^2=\d\mathbb{X}^2  &\qquad  {\cal P}^5=\d\mathbb{X}^5-mx\d\mathbb{X}^3-nx\d\mathbb{X}^6+mnx\mathbb{X}^2\d x   \\
{\cal P}^3=\d\mathbb{X}^3+nx\d\mathbb{X}^2 &\qquad {\cal P}^6=\d\mathbb{X}^6-m\mathbb{X}^2\d x
\end{array}\nonumber
\end{eqnarray}
In order to find a doubled torus bundle description, the $U(1)$ symmetry generated by $\widetilde{X}^m=\Pi^x{}_ML^{MN}\widetilde{\cal Z}_N=\widetilde{\cal Z}_4$ must be gauged in the sigma model (\ref{fibre action}). The vector field
$$
\widetilde{X}^x=\frac{\partial}{\partial {\ti x}}
$$
acts on ${\ti x}$ as ${\ti x}\rightarrow{\ti x}+\tilde{\varepsilon}_x$ and leaves all other coordinates invariant. From (\ref{generator transformations}) we can also see that this generator preserves and is preserved by $\G$ and therefore is well defined both on the doubled group $\cG$ and on the doubled twisted torus $\cX$. We expect to recover a five-dimensional doubled torus bundle ${\cal T}$ as the quotient
$$
{\cal T}=\cX/\widetilde{S}^1
$$
where $\widetilde{S}^1$ is the circle action generated by $\widetilde{X}^x$. Furthermore, since $\widetilde{X}^x$ preserves and is preserved by $\G$, the quotient $\cX/\widetilde{S}^1$ is well-defined and is a conventional Riemannian geometry.

As described in section three and also in \cite{New,Hull ``Doubled geometry and T-folds''}, we introduce world-sheet one-forms $C_x$ which transform as $\delta C_x=-\d\tilde{\varepsilon}_x$ so that $Q_x+C_x$ is invariant under the $U(1)$ gauge symmetry generated by $\widetilde{X}^x$. The kinetic term for the gauged sigma model may be written in terms of the Lagrangian
$$
{\cal L}_{\text{Kin}}=\frac{1}{4}P^x\wedge*P^x+\frac{1}{4}(Q_x+C_x)\wedge*(Q_x+C_x)+\frac{1}{4}{\cal M}_{AB}{\cal P}^A\wedge*{\cal P}^B
$$
It is useful to introduce the complement $\Pi^A{}_M$ of the partial polarisation such that
$$
{\cal P}^A=\Pi^A{}_M{\cal P}^M  \qquad  ({\cal M}^{-1})^{AB}=\Pi^A{}_M({\cal M}^{-1})^{MN}\Pi_N{}^B
$$
where ${\cal P}^A=({\cal P}^2,{\cal P}^3,{\cal P}^5,{\cal P}^6)$ and ${\cal M}_{AB}$ is a $4\times 4$ matrix, parameterising the coset $O(2,2)/O(2)\times O(2)$. The gauged Wess-Zumino term is
\begin{eqnarray}
S_{\text{c}}+S_{\text{wz}}&=&+\frac{1}{2}\oint_{\Sigma}P^x\wedge C_x-\frac{1}{2}\int_V mP^x\wedge{\cal P}^2\wedge{\cal P}^3-\frac{1}{2}\int_V nP^x\wedge{\cal P}^2\wedge{\cal P}^6\nonumber\\
&=&\frac{1}{2}\oint_{\Sigma}P^x\wedge (Q_x+C_x)\nonumber
\end{eqnarray}
The action for the $U(1)$-gauged sigma model is then given by
$$
S_{\cX/\widetilde{S}^1}=\oint_{\Sigma}{\cal L}_{\text{Kin}}+S_{\text{c}}+S_{\text{wz}}
$$
Completing the square in $C_x$, the action splits into two parts
$$
S_{\cX/\widetilde{S}^1}[C_x,x,{\ti x},\XX^A]=S_{\cal T}[x,\XX^A]+S_{\lambda}[C_x,{\ti x}]
$$
where
\begin{equation}\label{S_T}
S_{\cal T}[x,\XX^A]=\frac{1}{2}\oint_{\Sigma}P^x\wedge* P^x+\frac{1}{4}\oint_{\Sigma}{\cal M}_{AB}{\cal P}^A\wedge*{\cal P}^B
\end{equation}
is the sigma model for a world-sheet embedding into a doubled torus bundle \cite{Hull ``A geometry for non-geometric string backgrounds''} and
$$
S_{\lambda}[C_x,{\ti x}]=\frac{1}{2}\oint_{\Sigma}\lambda_x\wedge*\lambda_x  \qquad   \text{where}    \qquad  \lambda_x=C_x+Q_x-*P^x
$$
Integrating out the $C_x$ gives the sigma model for a world-sheet embedding into a doubled torus bundle ${\cal T}=\cG' / \G'$ where $\cG'$ is the five-dimensional Lie group $\cG / \widetilde{S}^1_x$ and $\G'$ is the discrete subgroup of $\G$ which does not act on ${\ti x}$. The five-dimensional Lie group $\cG'$ is a sub-group of a contraction of the six-dimensional Lie group $\cG$. This particular contraction is given by re-scaling the generator $\widetilde{X}^x\rightarrow \lambda \widetilde{W}^x$, taking the limit in which $\lambda\rightarrow 0$. The sub-algebra of this contracted algebra that generates $\cG'$ is given by the generators $\widetilde{Z}_x$ and $\widetilde{\cal Z}_A$, where the $\widetilde{\cal Z}_A$ are the vector fields  dual to the $\widetilde{\cal P}^A=\Pi^A{}_M\widetilde{\cal P}^M$.

In order to describe this five-dimensional geometry as a $T^4$ bundle over $S^1$, it is useful to redefine the coordinate
$$
\mathbb{X}^6 \rightarrow \mathbb{X}^6+mx\mathbb{X}^2
$$
so that the left-invariant one-forms in (\ref{S_T}) become
\begin{eqnarray}
\begin{array}{ll}
P^x=\d x  &   \\
{\cal P}^2=\d\mathbb{X}^2  &\qquad  {\cal P}^5=\d\mathbb{X}^5-mx\d\mathbb{X}^3-nx\d\mathbb{X}^6+mnx\mathbb{X}^2\d x   \\
{\cal P}^3=\d\mathbb{X}^3+nx\d\mathbb{X}^2 &\qquad {\cal P}^6=\d\mathbb{X}^6+mx\d\mathbb{X}^2
\end{array}\nonumber
\end{eqnarray}
This parameterisation of $\cal T$ is useful as these one-forms may then be written concisely as
$$
P^x=\d x  \qquad  {\cal P}^A=(e^{Nx})^A{}_B\d\mathbb{X}^B
$$
In this parameterisation, it is clear that the five-dimensional geometry is a $T^4$ bundle over $S^1$ with monodromy $e^{-N}$. Choosing the nilmanifold with $H$-flux polarisation, the monodromy of the fibre coordinates $\XX^A\sim (e^{-N})^A{}_B\XX^B$ can be written as
\begin{equation}\label{monodromy1}
e^{-N}={\cal O}_A \cdot{\cal O}_b=\left(
         \begin{array}{cc}
           A & 0 \\
           0 & (A^{-1})^T \\
         \end{array}
       \right)\left(
                \begin{array}{cc}
                  1 & b \\
                  0 & 1 \\
                \end{array}
              \right)
\end{equation}
where
$$
A=\left(
    \begin{array}{cc}
      1 & n \\
      0 & 1 \\
    \end{array}
  \right)   \qquad  b=\left(
                        \begin{array}{cc}
                          0 & m \\
                          -m & 0 \\
                        \end{array}
                      \right)
$$
Whereas in the twisted T-fold polarisation the monodromy is
\begin{equation}\label{monodromy2}
e^{-N}={\cal O}_A\cdot{\cal O}_{\beta}=\left(
         \begin{array}{cc}
           A & 0 \\
           0 & (A^{-1})^T \\
         \end{array}
       \right)\left(
                \begin{array}{cc}
                  1 & 0 \\
                  \beta & 1 \\
                \end{array}
              \right)
\end{equation}
where
$$
A=\left(
    \begin{array}{cc}
      1 & 0 \\
      -m & 1 \\
    \end{array}
  \right)   \qquad  \beta=\left(
                        \begin{array}{cc}
                          0 & n \\
                          -n & 0 \\
                        \end{array}
                      \right)
$$
Dualising along the $z$ direction simply has the effect of interchanging $m$ and $n$ in these monodromy matrices. It is interesting to note that, instead of gauging $U(1)_x$ generated by $\widetilde{X}^x$ and recovering the doubled torus bundle as ${\cal T}=\cX/\widetilde{S}^1_x$, we could have chosen the polarisation
$$
\Pi^y{}_M=\left(
            \begin{array}{cccccc}
              0 & 1 & 0 & 0 & 0 & 0 \\
            \end{array}
          \right)   \qquad  \widetilde{\Pi}_{yM}=\left(
            \begin{array}{cccccc}
              0 & 0 & 0 & 0 & 1 & 0 \\
            \end{array}
          \right)
$$
and gauged the $U(1)_y$ generated by $\widetilde{X}^y=\Pi^{yM}\widetilde{\cal Z}_M=\widetilde{\cal Z}_5$, which is well-defined on both $\cG$ and $\cX$. This gives a doubled torus bundle ${\cal T}'=\cX/\widetilde{S}^1_y$, where the $T^4$ fibres with coordinates $\XX^A=(\XX^1,\XX^3,\XX^4,\XX^6)$ are fibred over the circle $S^1_y$ with coordinate $y\sim y+1$ and monodromy $\XX^A\sim (e^{-N'})^A{}_B\XX^B$. How does this second doubled torus bundle, ${\cal T}'=\cX/\widetilde{S}^1_y$, differ from the first, ${\cal T}=\cX/\widetilde{S}^1_x$? The monodromy for this second doubled torus bundle is almost identical to (\ref{monodromy1}) and (\ref{monodromy2}) in the nilmanifold and T-fold polarisations respectively. The only difference is the sign change $(m,n)\rightarrow (-m,-n)$. This is to be expected as the set of left-invariant one-forms (\ref{left invariant forms}) which define the local structure of the doubled geometry are invariant under the exchange $(\XX^1,\XX^2,\XX^4,\XX^5,m,n)\leftrightarrow (\XX^2,\XX^1,\XX^5,\XX^4,-m,-n)$, which is equivalent to the exchange $(\Pi^x{}_M,\widetilde{\Pi}_{xM},m,n)\leftrightarrow (\Pi^y{}_M,\widetilde{\Pi}_{yM},-m,-n)$ once a maximally isotropic polarisation has been chosen.

\section{Inclusion of fields on $M_d$}

So far we have considered the sigma model with a target space given by the $2D$-dimensional doubled twisted torus $\cX$. This only describes the physics of the $D$-dimensional internal space of the lift of the gauged supergravity (\ref{D dim sugra}) to $(D+d)$-dimensions. In the specific examples we have considered, the scalar potential $V(\phi,{\cal M})$ in (\ref{D dim sugra}) vanishes if $m=n$ and so solutions of the equations of motion for $(\ref{D dim sugra})$ can be found where the $(D+d)$-dimensional space takes the form $M_d\times {\cal N}$; $M_d$ is flat, ${\cal N}$ is any of the three-dimensional backgrounds described in the last section, and the fields $H$ and ${\cal A}^M$ are both zero. In such limited cases, where the fields ${\cal A}^M$ and $H$ do not play a role, the description of the doubled sigma model with target space $\cX$ is adequate; however, a description of more interesting solutions to (\ref{D dim sugra}) requires us to describe the world-sheet embedding into the full $(D+d)$-dimensional space. In particular, a more complete description of the lift of the gauged supergravity to string theory must involve the fields on the $d$-dimensional base manifold $M_d$, including the connection one-forms ${\cal A}^M$ which describe the fibration of $\cX$ over $M_d$. The doubled world-sheet theory corresponding to the full gauged supergravity (\ref{D dim sugra}) may be thought of as a sigma model whose target space is a $(2D+d)$-dimensional bundle ${\cal Y}$
\begin{eqnarray}
\cX\hookrightarrow & {\cal Y} &\nonumber\\
&\downarrow&\nonumber\\
&M_d&\nonumber
\end{eqnarray}
with fibre given by the doubled twisted torus $\cX$ and base $M_d$. We choose local coordinates $X^{\cal I}=(z^{\mu},\XX^I)$, where $z^{\mu}$ are coordinates on the base, as in section one and $\XX^I$ are doubled coordinates on the fibre $\cX$. The connection for this bundle is ${\cal A}^M$ and we require that the transition functions for the bundle take values in the generalised T-duality group $\mathcal{O}(\Z)$. It is not clear what ${\cal O}(\Z)$ is for general $\cX$. We know that, when $\cX=T^{2D}$ then ${\cal O}(\Z)=O(D,D;\Z)$ and in \cite{New} it was suggested that the natural suggestion for the general form of ${\cal O}(\Z)$ should be ${\cal O}(\Z)=Aut(\cG;\G,L)$, the group of automorphisms
of $\cG$ that preserve $\G$ and the $O(D,D) $ metric $L_{MN}$. This gives the correct result when $\cG=\R^{2D}$, $\G=\Z^{2D}$ so that $\cG/\G=T^{2D}$ and $Aut(\cG;\G,L)=O(D,D;\Z)$.

The sigma model whose target space is the fibre $\cX$ is
given by the action (\ref{fibre action}).
If we define
$$
\widehat{\mathcal{P}}^M={\cal P}^M+{\cal A}^M
$$
then the action for the sigma model describing the embedding of $\Sigma$ into ${\cal Y}$ is
$$
S_{\cal Y}=\frac{1}{4}\oint_{\Sigma}{\cal M}_{MN}\widehat{\mathcal{P}}^M\wedge *\widehat{\mathcal{P}}^N +\frac{1}{2}\int_V
L_{MN}\widehat{\mathcal{P}}^M\wedge \mathcal{F}^N +\frac{1}{12}\int_Vt_{MNP}\widehat{\mathcal{P}}^M\wedge
\widehat{\mathcal{P}}^N\wedge \widehat{\mathcal{P}}^P+S[z]
$$
where the $S[z]$ is the action for the sigma model on the base $M_d$, given by
$$
S[z]=\frac{1}{2}\oint_{\Sigma}g_{\mu\nu}\d z^{\mu}\wedge *\d z^{\nu}+\int_VH
$$
The background fields appearing in $S_{\cal Y}$ are all pull-backs of the fields appearing in the gauged supergravity (\ref{D dim sugra}) to the
world-sheet. Using the identity
\begin{eqnarray}
&&\frac{1}{2}\int_V L_{MN}\widehat{\mathcal{P}}^M\wedge \mathcal{F}^N +\frac{1}{12}\int_Vt_{MNP}\widehat{\mathcal{P}}^M\wedge \widehat{\mathcal{P}}^N\wedge \widehat{\mathcal{P}}^P
-\frac{1}{2}\int_V\Omega_{\text{cs}}\nonumber\\&&=
-\frac{1}{2}\oint_{\Sigma}L_{MN}{\cal P}^M\wedge {\cal A}^N + \frac{1}{12}\int_Vt_{MNP}\mathcal{P}^M\wedge \mathcal{P}^N\wedge \mathcal{P}^P\nonumber
\end{eqnarray}
where $\Omega_{\text{cs}}$ is the pull-back of the Chern-Simons term (\ref{chern simons}) to the three-dimensional extension of the world-sheet $V$, the sigma-model action may be cast in a more convenient form
\begin{equation}\label{bundle action}
S_{\cal Y}=\frac{1}{4}\oint_{\Sigma}{\cal M}_{MN}\mathcal{P}^M\wedge *\mathcal{P}^N+\frac{1}{2}\oint_{\Sigma} \mathcal{P}^M\wedge
*{\cal I}_M+\frac{1}{12}\int_Vt_{MNP}\mathcal{P}^M\wedge \mathcal{P}^N\wedge \mathcal{P}^P+ S'[z]
\end{equation}
where
\begin{equation}\label{S'(Y)}
S'[z]= \frac{1}{2}\oint_{\Sigma}\left(g_{\mu\nu}+\frac{1}{2}{\cal M}_{MN}\mathcal{A}_{\mu}^M\mathcal{A}^N_{\nu}\right)\d z^{\mu}\wedge*\d z^{\nu}+\oint_{\Sigma}C_{(2)}
\end{equation}
where $C_{(2)}$ is related to the three-form $H$ by (\ref{C def}). It is useful to define
\begin{equation}\label{J definition}
{\cal I}_M={\cal M}_{MN}{\cal A}^N-L_{MN}*{\cal A}^N
\end{equation}

\subsection{Equations of motion and constraints}

The equation of motion for the scalar fields $\XX^I$, given by the action (\ref{bundle action}), is
$$
\d*{\cal M}_{MN}{\cal P}^N+M_{NP}t_{MQ}{}^P{\cal P}^Q\wedge *{\cal P}^N+L_{MN}\d{\cal P}^N=-\d*{\cal I}_M+t_{MP}{}^N{\cal P}^P\wedge *{\cal I}_N
$$
which, upon application of the Maurer-Cartan equations (\ref{worldsheet MC equations}) for ${\cal P}^M$, becomes
$$
\d\left({\cal M}_{MN}*{\cal P}^N+*{\cal I}_M+L_{MN}{\cal P}^N\right)+t_{MQ}{}^P{\cal P}^Q\wedge \left(M_{NP}*{\cal P}^N-*{\cal I}_P\right)=0
$$
Consider the self-duality constraint
\begin{equation}\label{constraint with spectators}
{\cal P}^M=L^{MN}\left({\cal M}_{NP}*{\cal P}^P+*{\cal I}_N\right)
\end{equation}
which is a generalisation of (\ref{constraintA}). Using this constraint and the Maurer-Cartan equations (\ref{worldsheet MC equations}), it is not hard to show that the equations of motion may be written as
$$
\d\left({\cal M}_{MN}*{\cal P}^N+*{\cal I}_M-L_{MN}{\cal P}^N\right)= 0
$$
and so the constraint (\ref{constraint with spectators}) is compatible with the equations of motion and the Maurer-Cartan equations. Repeated application of the self-duality constraint requires that ${\cal M}_{MN}$ satisfies (\ref{M=LML2}) and
\begin{equation}
{\cal I}_M=-L_{MN}{\cal M}^{NP}*{\cal I}_P
\end{equation}
This latter condition is identically satisfied from the definition of ${\cal I}_M$ in (\ref{J definition}). Writing out ${\cal I}_M$ in terms of the one-forms ${\cal A}^M$, the constraint (\ref{constraint with spectators}) may also be written as
$$
({\cal P}^M+{\cal A}^M)=L^{MN}{\cal M}_{NP}*({\cal P}^P+{\cal A}^P)
$$
which is equivalent to the constraints (\ref{constraintA}) and (\ref{constraint of A}).

\subsection{Choosing a polarisation and gauging the sigma model}

Once we choose a polarisation, the fibre bundle connection decomposes as
\begin{eqnarray}\label{A field}
\Pi^m{}_M{\cal A}^{M}{}_{\mu}= A^m{}_{\mu} \qquad  \widetilde{\Pi}_{mM}{\cal A}^M{}_{\mu}=B_{\mu m}
\end{eqnarray}
and we define the decomposition of ${\cal I}_M$ to be
\begin{eqnarray}
\Pi^{mM}{\cal I}_M=K^m    \qquad  \widetilde{\Pi}_m{}^M{\cal I}_M=J_m
\end{eqnarray}
where $J_m$ and $K^m$ are functions of the fields $A^m{}_{\mu}$ and $B_{\mu m}$. We see then that the polarisation not only selects which elements of ${\cal H}_{IJ}$ will be considered as metric degrees of freedom, it also selects which elements of ${\cal A}^M{}_{\mu}$ are to be selected as the connection for the fibration of the internal background over $M_d$. A similar construction was seen for the doubled torus bundle in \cite{Hull ``A geometry for non-geometric string backgrounds''}. The explicit form of these functions may be found by choosing
a polarisation and substituting (\ref{A field}) into (\ref{J definition}) which gives
$$
g_{mn}K^n=B_{\mu m}\d z^{\mu}-B_{mn}A^n{}_{\mu}\d z^{\mu}-g_{mn}A^n{}_{\mu}*\d z^{\mu}
$$
The expression for $J_m$ is more complicated and it is more useful to relate $J_m$ to $K^m$ by the relation ${\cal I}_M=-{\cal M}_{MN}L^{NP}*{\cal I}_P$ which
explicitly gives
$$
J_m=-g_{mn}*K^n+B_{mn}K^n
$$
As discussed in previous sections, the polarisation selects a sub-group $\widetilde{G}_L$ of $\cG_L$ which can be gauged to give a sigma model which describes the embedding of the world-sheet into (a patch of) the quotient ${\cal Y}/\widetilde{G}_L$. Here we shall assume that the gauging can be done in each fibre $\cX$ (or an appropriate cover of $\cX$ when the polarisation selects a T-fold for the internal geometry as in section 4.2) and we will not be concerned here about the global structure of how the internal background is fibred over $M_d$. The details of the global structure of this quotient, where $M_d$ may contain non-contractible cycles, will be briefly commented on in section 6.4. Introducing the world-sheet one-forms $C_m=C_{m\alpha}\d\sigma^{\alpha}$ as in section 3.2.2, and the $\cG$-twisted world-sheet one-forms ${\cal C}=g^{-1}Cg$, where $g\in\cG$, the gauged sigma model may be written as
\begin{eqnarray}
S_{{\cal Y}/\widetilde{G}_L}&=&\frac{1}{4}\oint_{\Sigma}{\cal M}_{MN}({\cal P}^M+{\cal C}^M)\wedge *({\cal P}^N+{\cal C}^N) +\frac{1}{2}\oint_{\Sigma}({\cal P}^M+{\cal C}^M)\wedge*{\cal I}_M\nonumber\\
 &&+\frac{1}{2}\oint_{\Sigma}L_{MN}{\cal P}^M\wedge{\cal C}^N +\frac{1}{12}\int_Vt_{MNP}{\cal P}^M\wedge{\cal P}^N\wedge {\cal P}^P+S'[z]\nonumber
\end{eqnarray}
Introducing the current ${\cal S}_M={\cal J}_M+{\cal I}_M$, which may be written as
$$
{\cal S}_M={\cal M}_{MN}\widehat{{\cal P}}^N-L_{MN}*\widehat{{\cal P}}^N
$$
this gauged action may be given by
\begin{eqnarray}\label{gauged lagrangian with spectators}
S_{{\cal Y}/\widetilde{G}_L}&=&\frac{1}{4}\oint_{\Sigma}{\cal M}_{MN}{\cal P}^M\wedge *{\cal P}^N +\frac{1}{2}\oint_{\Sigma}{\cal P}^M\wedge*{\cal I}_M +\frac{1}{12}\int_Vt_{MNP}{\cal P}^M\wedge{\cal P}^N\wedge {\cal P}^P\nonumber\\
 &&+\frac{1}{2}\oint_{\Sigma}{\cal C}^M\wedge*{\cal S}_M +\frac{1}{4}\oint_{\Sigma}{\cal M}_{MN}{\cal C}^M\wedge*{\cal C}^N +S'[z]
\end{eqnarray}
In particular, the gauging imposes the constraint ${\cal S}_M=0$, which is equivalent to the self-duality constraint (\ref{constraint with spectators}).

\subsection{Recovering the flux compactification on a twisted torus}

In this section we demonstrate how a sigma model for the flux compactification on a $D$-dimensional twisted torus discussed in the Introduction (see also \cite{Hull ``Flux compactifications of string theory on twisted tori'',Kaloper ``The O(dd) story of massive supergravity''}) may be recovered from the doubled sigma model (\ref{gauged lagrangian with spectators}) with target space ${\cal Y}$ by a judicious choice of polarisation. This example is based on a doubled group $\cG$ which may be thought of as a generalisation of the six-dimensional doubled group considered in sections three and four. We select a polarisation in which the only non-vanishing structure constants are
$$
\widetilde{\Pi}_m{}^M\widetilde{\Pi}_n{}^N\Pi^p{}_Pt_{MN}{}^P=f_{mn}{}^p    \qquad  \widetilde{\Pi}_m{}^M\widetilde{\Pi}_n{}^N\widetilde{\Pi}_{pP}t_{MN}{}^P=K_{mnp}
$$
The nilmanifold with constant $H$-flux is a particular example of such a polarisation for such a doubled group. In this polarisation, the
Maurer-Cartan equations (\ref{worldsheet MC equations}) are
$$
\d P^m+\frac{1}{2}f_{np}{}^mP^n\wedge P^p=0   \qquad  \d Q_m- f_{mn}{}^pQ_p\wedge P^n-\frac{1}{2}K_{mnp}P^n\wedge P^p=0
$$
and the Wess-Zumino term may be written as
\begin{eqnarray}
S_{\text{wz}}&=&\frac{1}{4}\int_Vf_{np}{}^mQ_m\wedge P^n\wedge P^p-\frac{1}{12}\int_VK_{mnp}P^m\wedge P^n\wedge P^p\nonumber\\
&=&\frac{1}{2}\oint_{\Sigma}P^m\wedge Q_m+\frac{1}{6}\int_VK_{mnp}P^m\wedge P^n\wedge P^p\nonumber
\end{eqnarray}
We must also consider the term which includes a coupling to the directions along the base of the fibration. In this polarisation, the term
becomes
\begin{eqnarray}
\frac{1}{2}\oint_{\Sigma}{\cal P}^{\hat{M}}\wedge *{\cal I}_{\hat{M}}&=&\frac{1}{2}\oint_{\Sigma}P^m\wedge *J_m+\frac{1}{2}\oint_{\Sigma}Q_m\wedge
*K^m\nonumber\\
&=&-\frac{1}{2}\oint_{\Sigma}g_{mn}P^m\wedge K^n +\frac{1}{2}\oint_{\Sigma}B_{mn}P^m\wedge *K^n +\frac{1}{2}\oint_{\Sigma}Q_m\wedge *K^m\nonumber
\end{eqnarray}
and the term which couples to the constraint current ${\cal S}_M$ is
\begin{eqnarray}
\frac{1}{2}\oint_{\Sigma}{\cal C}^{\hat{M}}\wedge *{\cal S}_{\hat{M}}&=&\frac{1}{2}\oint_{\Sigma}\mathcal{C}_m\wedge (-g^{mp}B_{pn}P^n +g^{mn}Q_n -*P^m+K^m)\nonumber
\end{eqnarray}
The gauged sigma model (\ref{gauged lagrangian with spectators}) then gives
\begin{eqnarray}
S_{{\cal Y}/\widetilde{G}_L}&=&\frac{1}{4}\oint_{\Sigma}g^{mn}(\mathcal{C}_m+Q_m)\wedge*(\mathcal{C}_n+Q_n) +\frac{1}{2}\oint_{\Sigma}(\mathcal{C}_m+Q_m)\wedge*(-g^{mp}B_{pn}P^n+K^m-*P^m)\nonumber\\
 &&+\frac{1}{4}\oint_{\Sigma}(g_{mn}-B_{mp}g^{pq}B_{qn})P^m\wedge*P^n +\frac{1}{6}\int_{V}K_{mnp}P^m\wedge P^n\wedge P^p \nonumber\\
 && +\frac{1}{2}\oint_{\Sigma}B_{mn}P^m\wedge*K^n -\frac{1}{2}\oint_{\Sigma}g_{mn}P^m\wedge K^n+S'[z]\nonumber
\end{eqnarray}
Completing the square in $\mathcal{C}_m+Q_m$, the action splits into two distinct parts
$$
S_{{\cal Y}/\widetilde{G}_L}[z,\XX,C]=S_Y[z,x]+S_{\lambda}[C,z,{\ti x}]
$$
where
\begin{eqnarray}
S_Y&=&\frac{1}{2}\oint_{\Sigma}g_{mn}P^m\wedge*P^n
 +\frac{1}{2}\oint_{\Sigma}B_{mn}P^m\wedge P^n -\oint_{\Sigma}g_{mn}P^m\wedge K^n-\frac{1}{4}\oint_{\Sigma}g_{mn}K^m\wedge*K^n \nonumber\\
  &&+\frac{1}{6}\int_{V}K_{mnp}P^m\wedge P^n\wedge P^p+S'[z]\nonumber
\end{eqnarray}
and
$$
S_{\lambda}=\frac{1}{4}\oint_{\Sigma}g^{mn}\lambda_m\wedge*\lambda_n
$$
where
\begin{eqnarray}
\lambda_m&=&{\cal C}_m+Q_m-g_{mn}*P^n-B_{mn}P^n+g_{mn}K^n\nonumber\\
&=&\mathcal{C}_m+(Q_m+B_{(1)m})-g_{mn}*\nu^m-B_{mn}\nu^n
\end{eqnarray}
The world-sheet one-forms $\nu^m=\nu^m{}_{\alpha}\d\sigma^{\alpha}$, given by
$$
\nu^m{}_{\alpha}=P^m{}_{\alpha}+A^m{}_{\alpha}
$$
 are the pull-back of the one-forms (\ref{nu}) to the world-sheet. Recalling that the action $S'(z)$ is given by (\ref{S'(Y)}) where the two-form $C_{(2)}$ is given by (\ref{C def}), the action $S_Y$ may be simplified further
\begin{eqnarray}
S_Y&=&\frac{1}{2}\oint_{\Sigma}g_{\mu\nu}\d z^{\mu}\wedge*\d z^{\nu} +\frac{1}{2}\oint_{\Sigma}g_{mn}\nu^m\wedge*\nu^n
 +\oint_{\Sigma}\left(B_{(2)}+B_{\mu m}\d z^{\mu}\wedge\nu^m +\frac{1}{2}B_{mn}\nu^m\wedge \nu^n\right) \nonumber\\
 && +\frac{1}{6}\int_{V}K_{mnp}P^m\wedge P^n\wedge P^p\nonumber
\end{eqnarray}
The $\lambda_m$ in $S_{\lambda}$ may be integrated out and give a shift in the dilaton contribution as described in \cite{New,Hull ``Doubled geometry and T-folds''}. Introducing coordinates $X^{\mathcal{I}}=(z^{\mu},x^i)$ on the $(D+d)$-dimensional bundle, the full action in the bundle may be written in the more conventional form
$$
S=\frac{1}{2}\oint_{\Sigma}\widehat{G}_{\mathcal{I}\mathcal{J}}\d X^{\mathcal{I}}\wedge*\d X^{\mathcal{J}} +\frac{1}{2}\oint_{\Sigma}\widehat{B}_{\mathcal{I}\mathcal{J}}\d X^{\mathcal{I}}\wedge \d X^{\mathcal{J}}
$$
where the metric and $B$-field are given by (\ref{reduction ansatz}) and the $H$-field strength is given by (\ref{H-flux}). We see that the sigma model (\ref{gauged lagrangian with spectators}) correctly recovers a sigma model on the twisted torus with constant $H$-flux described in \cite{Hull ``Flux compactifications of string theory on twisted tori'',Kaloper ``The O(dd) story of massive supergravity''} and reviewed in the Introduction.

\subsection{Global issues and generalised T-folds}

In \cite{Hull ``A geometry for non-geometric string backgrounds''}, a T-fold was defined to be a background which is locally a Riemannian geometry, but is globally patched together using transition functions which include strict T-dualities. The canonical example is of an $n$-dimensional torus fibration in which the transition functions take values in the non-geometric elements of $O(n,n;\Z)$. The sigma model describing a world-sheet embedding into the bundle ${\cal Y}$ studied in this section suggests a generalisation of this T-fold construction, to what might be called a generalised T-fold. We define a generalised T-fold to be a background which is patched together by transition functions which may include generalised (possibly non-isometric) T-dualities, i.e. an element of ${\cal O}(\Z)$. Bundles ${\cal Y}$, where $M_d$ has non-contractible cycles over which $\cX$ is non-trivially fibred, may provide examples of such generalised T-folds. In particular, if $M_d=S^1$ and the monodromy of the theory in the fibres $\cX$ includes a non-geometric element of ${\cal O}(\Z)$ which is not a strict T-duality of the kind described by the Buscher rules, but is a generalised T-duality of the kind conjectured in \cite{Dabholkar ``Generalised T-duality and non-geometric backgrounds''}, then the fibration $Y$ is a generalised T-fold. In this case a generalised T-fold can be constructed if we can identify two backgrounds, related by a generalised T-duality, which are also homotopic to each other. This is a particular suggestion for such a background but more general constructions involving more general $M_d$ may also be possible. It would be interesting to see if the $K3$ mirror-folds constructed in \cite{ReidEdwards:2008rd,Kawai Mirrorfolds} can be understood in this way. The key issue in all such constructions is to ensure that the background is patched together by transition functions which act as true symmetries of the string theory, then these generalised T-folds will be candidates for smooth string theory backgrounds. The crucial issue then is to identify the group ${\cal O}(\Z)$ in interesting examples and in particular, to determine whether or not it coincides with $Aut(\cG;\G,L)$, as suggested in \cite{New}.

\begin{center}
\textbf{Acknowledgements}
\end{center}

\noindent I am grateful to the Centre for Research in String Theory at Queen Mary, University of London for their continued kind hospitality.

\begin{footnotesize}

\end{footnotesize}

\end{document}